\documentclass[twocolumn]{aastex631}

\shorttitle{
}
\shortauthors{Hatano et al.}
\graphicspath{{./}{figures/}}
\usepackage[utf8]{inputenc}   
\usepackage{amssymb, amsmath}

\usepackage{ulem}
\usepackage{xcolor}
\DeclareRobustCommand{\erase}{\bgroup\markoverwith{\textcolor{red}{\rule[.5ex]{2pt}{0.4pt}}}\ULon}

\date{Accepted 7 March, 2024}
\defcitealias{2022ApJ...930...37U}{U22}

\begin{document}
\author[0000-0002-5816-4660]{Shun Hatano}
\affiliation{National Astronomical Observatory of Japan, Osawa 2-21-1, Mitaka, Tokyo 181-8588, Japan}
\affiliation{Department of Astronomical Science, The Graduate University for Advanced Studies, SOKENDAI, 2-21-1 Osawa, Mitaka, Tokyo, 181-8588, Japan}

\author[0000-0002-1049-6658]{Masami Ouchi}
\affiliation{National Astronomical Observatory of Japan, Osawa 2-21-1, Mitaka, Tokyo 181-8588, Japan}
\affiliation{Institute for Cosmic Ray Research, The University of Tokyo, 5-1-5 Kashiwanoha, Kashiwa, Chiba 277-8582, Japan}
\affiliation{Astronomical Science Program, Graduate Institute for Advanced Studies, SOKENDAI, 2-21-1 Osawa, Mitaka, Tokyo 181-8588, Japan}
\affiliation{Kavli Institute for the Physics and Mathematics of the Universe (WPI), University of Tokyo, Kashiwa, Chiba 277-8583, Japan}

\author{Hiroya Umeda}
\affiliation{Institute for Cosmic Ray Research, The University of Tokyo, 5-1-5 Kashiwanoha, Kashiwa, Chiba 277-8582, Japan}
\affiliation{Department of Physics, Graduate School of Science, The University of Tokyo, 7-3-1 Hongo, Bunkyo, Tokyo 113-0033, Japan}

\author[0000-0003-2965-5070]{Kimihiko Nakajima}
\affiliation{National Astronomical Observatory of Japan, Osawa 2-21-1, Mitaka, Tokyo 181-8588, Japan}

\author[0000-0002-3866-9645]{Toshihiro Kawaguchi}
\affiliation{Department of Economics, Management and Information Science,
  Onomichi City University, Hisayamada 1600-2, Onomichi, Hiroshima
722-8506, Japan}

\author[0000-0001-7730-8634]{Yuki Isobe}
\affiliation{Institute for Cosmic Ray Research, The University of Tokyo, 5-1-5 Kashiwanoha, Kashiwa, Chiba 277-8582, Japan}
\affiliation{Department of Physics, Graduate School of Science, The University of Tokyo, 7-3-1 Hongo, Bunkyo, Tokyo 113-0033, Japan}

\author[0000-0002-1005-4120]{Shohei Aoyama}
\affiliation{Institute of Management and Information Technologies, Chiba University, 1-33, Yayoi-cho, Inage-ward, Chiba, 263-8522, Japan}
\affiliation{Institute for Cosmic Ray Research, The University of Tokyo, 5-1-5 Kashiwanoha, Kashiwa, Chiba 277-8582, Japan}

\author{Kuria Watanabe}
\affiliation{National Astronomical Observatory of Japan, Osawa 2-21-1, Mitaka, Tokyo 181-8588, Japan}
\affiliation{Department of Astronomical Science, The Graduate University for Advanced Studies, SOKENDAI, 2-21-1 Osawa, Mitaka, Tokyo, 181-8588, Japan}

\author[0000-0002-6047-430X]{Yuichi Harikane}
\affiliation{Institute for Cosmic Ray Research, The University of Tokyo, 5-1-5 Kashiwanoha, Kashiwa, Chiba 277-8582, Japan}

\author[0000-0002-3801-434X]{Haruka Kusakabe} 
\affiliation{Observatoire de Gen{\'e}ve, Universit{\'e} de Gen{\'e}ve, 51 Ch. des Maillettes, 1290 Versoix, Switzerland}
\affiliation{National Astronomical Observatory of Japan, Osawa 2-21-1, Mitaka, Tokyo 181-8588, Japan}

\author{Akinori Matsumoto}
\affiliation{Institute for Cosmic Ray Research, The University of Tokyo, 5-1-5 Kashiwanoha, Kashiwa, Chiba 277-8582, Japan}
\affiliation{Department of Physics, Graduate School of Science, The University of Tokyo, 7-3-1 Hongo, Bunkyo, Tokyo 113-0033, Japan}
\author[0000-0003-1169-1954]{Takashi J. Moriya}
\affiliation{National Astronomical Observatory of Japan, Osawa 2-21-1, Mitaka, Tokyo 181-8588, Japan}
\affiliation{School of Physics and Astronomy, Faculty of Science, Monash University, Clayton, Victoria 3800, Australia}
\author[0000-0003-4321-0975]{Moka Nishigaki}
\affiliation{National Astronomical Observatory of Japan, Osawa 2-21-1, Mitaka, Tokyo 181-8588, Japan}
\affiliation{Department of Astronomical Science, The Graduate University for Advanced Studies, SOKENDAI, 2-21-1 Osawa, Mitaka, Tokyo, 181-8588, Japan}

\author[0000-0001-9011-7605]{Yoshiaki Ono}
\affiliation{Institute for Cosmic Ray Research, The University of Tokyo, 5-1-5 Kashiwanoha, Kashiwa, Chiba 277-8582, Japan}

\author[0000-0003-3228-7264]{Masato Onodera}
\affiliation{Subaru Telescope, National Astronomical Observatory of Japan, National Institutes of Natural Sciences (NINS), 650 North A'ohoku Place, Hilo, HI 96720, USA}

\author[0000-0001-6958-7856]{Yuma Sugahara} 
\affiliation{National Astronomical Observatory of Japan, Osawa 2-21-1, Mitaka, Tokyo 181-8588, Japan}
\affiliation{Waseda Research Institute for Science and Engineering, Faculty of Science and Engineering, Waseda University, 3-4-1, Okubo, Shinjuku, Tokyo 169-8555, Japan}

\author[0000-0002-7043-6112]{Akihiro Suzuki}
\affiliation{Research Center for the Early Universe, The University of Tokyo, 7-3-1 Hongo, Bunkyo, Tokyo 113-0033, Japan}

\author[0000-0002-5768-8235]{Yi Xu}
\affiliation{Institute for Cosmic Ray Research, The University of Tokyo, 5-1-5 Kashiwanoha, Kashiwa, Chiba 277-8582, Japan}
\affiliation{Department of Astronomy, Graduate School of Science, the University of Tokyo, 7-3-1 Hongo, Bunkyo, Tokyo 113-0033, Japan}

\author[0000-0003-3817-8739]{Yechi Zhang}
\affiliation{Institute for Cosmic Ray Research, The University of Tokyo, 5-1-5 Kashiwanoha, Kashiwa, Chiba 277-8582, Japan}
\affiliation{Department of Astronomy, Graduate School of Science, the University of Tokyo, 7-3-1 Hongo, Bunkyo, Tokyo 113-0033, Japan}
\affiliation{Kavli Institute for the Physics and Mathematics of the Universe (WPI), University of Tokyo, Kashiwa, Chiba 277-8583, Japan}

\title{
EMPRESS. XIV.
Strong High Ionization Lines of Young Galaxies at $z=0-8$:\\
Ionizing Spectra Consistent with the Intermediate Mass Black Holes with $M_{\rm BH}\sim 10^3-10^6\ M_\odot$}

\begin{abstract}

We present ionizing spectra estimated at 13.6--100 eV for ten dwarf galaxies with strong high ionization lines of He {\sc {ii}}$\lambda$4686 and [Ne {\sc{v}}]$\lambda$3426 ([Ne {\sc{iv}}]$\lambda$2424) at $z=0$ ($z=8$) that are identified in our Keck/LRIS spectroscopy and the literature (the JWST ERO program). With the flux ratios of these high ionization lines and $>10$ low-ionization lines of hydrogen, helium, oxygen, neon, and sulfur, we determine ionizing spectra consisting of stellar and non-thermal power-law radiation by photoionization modeling with free parameters of nebular properties including metallicity and ionization parameter, cancelling out abundance ratio differences. We find that all of the observed flux ratios are well reproduced by the photoinization models with the power law index $\alpha_{\rm EUV}$ of $\alpha_{\rm EUV}\sim (-1)-0$ and the luminosity $L_{\rm EUV}$ of $L_{\rm EUV}\sim 10^{40}-10^{42}$ erg s$^{-1}$ at $\sim 55-100$ eV for six galaxies, while four galaxies include large systematics in $\alpha_{\rm EUV}$ caused by stellar radiation contamination. We then compare $\alpha_{\rm EUV}$ and $L_{\rm EUV}$ of these six galaxies with those predicted by the black hole (BH) accretion disk models, and find that $\alpha_{\rm EUV}$ and $L_{\rm EUV}$ are similar to those of the intermediate mass black holes (IMBHs) in BH accretion disk models {albeit with possibilities of the other scenarios.} Confirming these results with a known IMBH having a mass $M_{\rm BH}$ of $M_{\rm BH}=10^{5.75} \ M_\odot$, we find that four local galaxies and one $z=7.665$ galaxy have ionizing spectra consistent with those of IMBHs with $M_{\rm BH} \sim 10^3-10^5 \ M_\odot$.
\end{abstract}

\section{Introduction}
Studies over the last two decades have revealed that massive galaxies are typically harboring super massive black holes (SMBH) with masses of $10^{6-10} M_\odot$ at their centers.
However, the formation of SMBH is puzzling. 
SMBHs with masses of $\sim 10^9\ M_\odot$ were already formed at $z\sim 6-7$ when the universe was 
equal to or less than 1 Gyr old (e.g., 
\citealt{2011Natur.474..616M, 2018Natur.553..473B, 2021ApJ...907L...1W}).
Because a spherical mass accretion onto a
BH is not very efficient due to
radiation pressure from accreting gas,
it is suggested that SMBHs form via
massive seed BHs with intermediate masses of 
$\sim 10^{2-5}\ M_\odot$ produced via population III stars 
and direct collapse (e.g., \citealt{2014ApJ...781...60H,2001ApJ...546..635O}).

Although such IMBHs ($M_{\rm BH} \sim 10^{2-5}\ M_\odot$) are not well 
understood in the high redshift as well as in the local universe
due to diffuculties of observations, 
recent optical surveys in the local universe have identified low-mass AGNs in dwarf galaxies that harbour BHs with masses down to $M_{\rm{BH}} \sim 10^5 \ M_\odot$ \citep{2011ApJ...739...28X}.
{There are growing number of IMBH candidates in dwarf galaxies with report of high-ionization lines or H$\alpha$ broad lines \citep{2018ApJ...863....1C,2021ApJ...922..155M,2022ApJ...936..140R, 2020MNRAS.494..941S, 2021ApJ...911...70B, 2021ApJ...912L...2C, 2021ApJ...922..170B, 2024MNRAS.tmp..271M}.}
On the other hand, gravitational wave observations have revealed IMBHs as well as stellar BHs with masses up to $M_{\rm{BH}} \sim 10^2 \ M_\odot$ \citep{2020ApJ...900L..13A}. 
There is a gap of BH masses known to date in the mass range of $M_{\rm{BH}}\sim 10^3 - 10^4\ M_\odot$,
which is a missing piece of SMBH formation \citep{2020ARA&A..58..257G}.

Among dwarf galaxies, extremely metal poor galaxies (EMPGs), 
are promising galaxies to explore the missing IMBHs with the masses of $M_{\rm{BH}}\sim 10^3 - 10^4\ M_\odot$. 
EMPGs are low metallicity $Z$ and stellar-mass $M_*$ galaxies, typically having 
$Z\sim 0.01-0.1\ Z_\odot$ and $10^5-10^8\ M_\odot$, respectively. The properties such as low metallicity and 
low stellar mass indicate the possibility that EMPGs are experiencing the early phase of star and BH formation.
Observations have shown the existence of hard ionizing radiation in EMPGs whose strong high ionization lines of He {\sc ii}$\lambda$4686 and {[Ne {\sc {v}}]$\lambda$3426} cannot be explained with 
stellar synthesis models alone 
{(e.g., \citealt{Izotov2004, 2005ApJS..161..240T, 2012MNRAS.427.1229I, 2021MNRAS.508.2556I, Schaerer+19}).}
{The He {\sc ii}$\lambda$4686 and [Ne {\sc v}]$\lambda$3426 lines are observed in various local galaxies including EMPG \citep{Garnett+91,Guseva+00,2005ApJS..161..240T,Brinchmann+08, 2010A&A...516A.104L, Shirazi+12, Izotov2004, 2012MNRAS.427.1229I, 2021ApJ...922..170B, 2021MNRAS.508.2556I}. }
{Fast radiative shocks, which may be produced by supernovae (SNe), are proposed to explain the observed [Ne {\sc v}]$\lambda$3426 emission line fluxes and/or ratios in some low-metallicity galaxies (e.g. \citealt{2005ApJS..161..240T, 2012MNRAS.427.1229I,2021MNRAS.508.2556I}). 
If the [Ne {\sc v}]$\lambda$3426 emission lines originate from SNe, the [Ne {\sc v}]$\lambda$3426 emission line fluxes may show temporal variability. 
We have searched for low-metallicity dwarf galaxies with multiple observations of [Ne {\sc v}]$\lambda$3426 lines and have found a galaxy dubbed SBS 0335-052E. 
The [Ne {\sc v}]$\lambda$3426 emission lines of SBS 0335-052E are observed in September 2003 and February 2004 \citep{Izotov2009, 2005ApJS..161..240T}. 
The extinction corrected emission line ratios of $\lbrack$Ne\ {\sc{v}}]$\lambda$3426/[Ne\ {\sc{iii}}]$\lambda$3869 are 0.0387 $\pm$ 0.0068 and 0.0306 $\pm$ 0.0026 for each observation. We find that the two ratios are consistent within $\sim$ 1-2 sigma, and we find no significant temporal variability in the [Ne\ {\sc{v}}]$\lambda$3426 lines.}
The hard ionizing radiation may be caused by BHs
(e.g., \citealt{2022ApJ...930...37U, 2021ApJ...908...68O, 2021A&A...656A.127S}).
{There is a possibility that X-ray binaries (XRBs) 
having the stellar-mass BHs 
produce the hard ionizing radiation \citep{2024ApJ...960...13G,Schaerer+19}, 
although a famous dwarf galaxy SBS 0335-052E, which shows strong He {\sc ii}$\lambda$4686 emission, has an X-ray luminosity lower than expected from XRBs.
(\citealt{2018MNRAS.480.1081K};
\citealt{2020MNRAS.496.3796S};
cf. \citealt{2004ApJ...606..213T}).
{ULXs have been proposed to explain for the observed [Ne {\sc {v}}]$\lambda$3426 for some dwarf galaxies \citep{2022ApJ...930...37U, 2021A&A...656A.127S}. }
Another possibility is ionizing radiation originating from a massive BH (e.g., IMBH or SMBH) residing at the center of a galaxy that produces X-ray spectum softer than that of a stellar-mass BH {\citep{2003ApJ...593...69K, 2005ApJS..161..240T, 2012MNRAS.427.1229I, Plat+19,2021MNRAS.508.2556I}.}
It should be noted that stellar masses of EMPGs are small, $M_* = 10^5 - 10^8\ M_\odot$
\citep{2021ApJ...918...54I,2022ApJ...925..111I,2020ApJ...898..142K,2021ApJ...922..170B,2016ApJ...819..110S}.
If the local stellar-mass to SMBH mass relation is extrapolated to the low mass regime, the BH masses may be about 1/1000 of the stellar masses, $M_{\rm BH} \sim 10^2 - 10^5\ M_\odot$ that fall in the missing IMBH mass range \citep{2015ApJ...813...82R}.
Theoretical models suggest that IMBHs have hot accretion disks with temperature of  $\sim 10^{5-6}$ K producing hard blackbody radiation in the low-energy X-ray band of 13.6 -- 100 eV \citep{2003ApJ...593...69K}.
However, such hard radiation in 13.6 -- 100 eV cannot be directly observed due to the strong hydrogen absorption around BHs.}

Here we estimate spectra of dwarf galaxies including EMPGs in the hard energy band, exploiting the photoionization modeling with observed optical emission line fluxes that is established by \citeauthor{2022ApJ...930...37U} (\citeyear{2022ApJ...930...37U}; hereafter \citetalias{2022ApJ...930...37U}).
\citetalias{2022ApJ...930...37U} reconstruct spectra in the energy range of 13.6--54 eV with observed emission lines from hydrogen Balmer lines to He{\sc ii}$\lambda 4686$ whose ionization energies range in 13.6--54 eV.
In this study, we extend the \citetalias{2022ApJ...930...37U} modeling technique by incorporating [Ne {\sc{v}}]$\lambda 3426$ and [Ne {\sc{iv}}]$\lambda2424$ emission lines, which have much higher ionizing energies of 97 eV and 63 eV, respectively, to cover the ionizing spectra up to $\sim 100$ eV and test the IMBH scenario,
{while we do not aim to rule out the other possibilities (e.g., HMXB, radiative shocks).}
Furthermore, the model of stellar contributions at $<60$ eV is refined compared to the \citetalias{2022ApJ...930...37U} model by using stellar synthesis model instead of simple blackbody radiation used in \citetalias{2022ApJ...930...37U} model, allowing for more accurate determination of the spectral shape of the hard component in the 55--100 eV band that is hereafter referred to as the extreme ultraviolet (EUV) band. 
{Because we free spectral index and amplitude of the power-law component, 
our SEDs are generalized and reproduce both soft and hard spectral shapes. }


{There are other ionizing spectra modeling techniques developed by 
\cite{2021ApJ...908...68O},\cite{2021A&A...656A.127S}, \cite{2024ApJ...960...13G}, and \cite{2021MNRAS.508.2556I}. 
\cite{2021ApJ...908...68O} use the sum of stellar synthesis models and blackbody radiation.
\cite{2021A&A...656A.127S} utilize observationally constrained spectra of ULXs and BPASS binary models. 
\cite{2024ApJ...960...13G} utilize the sum of BPASS models and XRB spectra. 
\cite{2021MNRAS.508.2556I} use non-thermal power-law radiation, BPASS single models, and STARBURST99 models.
{Because our spectra free the spectra index and amplitude of power-law component,}
the SEDs of our models include those of \cite{2021ApJ...908...68O}, \cite{2021A&A...656A.127S}, \cite{2024ApJ...960...13G}, and \cite{2021MNRAS.508.2556I}.
}

{
It is suggested that temperature and electron density structures are dependent on ionizing spectra \citep{2021MNRAS.508..680B}. 
To include such conditions, some studies utilize multiple-zone models (e.g. \citealt{2021ApJ...922..170B}).
However, it is unknown whether all the emission lines including H$\beta$, He {\sc ii}$\lambda$4686, and [Ne {\sc v}]$\lambda$3426 can be reproduced in a one-zone model. 
In fact, \cite{2021MNRAS.508.2556I} do not find a solution of a one-zone model reproducing He {\sc ii}$\lambda$4686/H$\beta$ and [Ne {\sc v}]$\lambda$3426/H$\beta$ ratios observed in some dwarf galaxies with [Ne {\sc v}]$\lambda$3426 detection (e.g. Tol 1214). 
In this study, we explore one-zone models that can reproduce emission line ratios. In this sense, multiple zone models are out of the scope in our study.
}

The structure of this paper is as follows.
In Section \ref{sec:observation}, we explain our Keck observations for EMPGs. We present our sample in Section \ref{sec:sample}. In Section \ref{sec:modeling}, we describe our modeling. In Section \ref{sec:result}, we apply our modeling technique to galaxies in our sample, and determine the best-estimate spectra over $\sim 13.6-100$ eV. We compare our best-estimate spectra with BH accretion disk models in section \ref{sec:discussion}. 
Throughout this paper, the magnitudes are in AB system, and we adopt a cosmological model with $H_0=70\ {\rm km \ s^{-1} \ Mpc^{-1}}$, $\Omega_{\rm \Lambda}=0.7$, and $\Omega_{\rm m} = 0.3$.
We use the solar metallicity scale of \cite{2009ARA&A..47..481A}, where 12 + log(O/H) = 8.69.

\section{Observation and data reduction}
\label{sec:observation}

\subsection{Spectroscopic Observations with Keck/LRIS}

We observed EMPGs including 2 bright famous EMPGs, SBS 0335-052E and HS 0122+0743 (hereafter HS 0122), with Keck/LRIS on 2021 November 7 and 8 (PI: K. Nakajima). 
We placed $0\farcs7$ wide slits for all of the targets. 
For the blue (red) channel of LRIS, we utilized the 600/4000 grism (600/7500 grating) which provides a spectral resolution of $\sim 4 \ (5)$ \AA\ in FWHM. 
The LRIS spectroscopy of the blue and red channels cover the wavelength ranges of $\lambda \sim 3000-5500$ and $6000-9000$ \AA, respectively. 
We also observed Feige 34 for flux calibration. 
The weather was clear during the observations, and 
the seeing sizes range in $0.8$ -- $1.0$ arcsec.
The spectroscopy of SBS 0335-052E taken in the Keck/LRIS observations is shown in \cite{2023arXiv230403726H}. Because they only show the spectra taken with the red arm of Keck/LRIS and the H$\alpha$ emission line fluxes of SBS 0335-052E, we present the spectra of the blue arm and the obtained fluxes and flux ratios of emission lines besides the H$\alpha$ emission line in this paper.

\subsection{Data Reduction }
\label{sec:data_reduction}
\begin{figure}
    \centering
    \includegraphics[width=80mm]{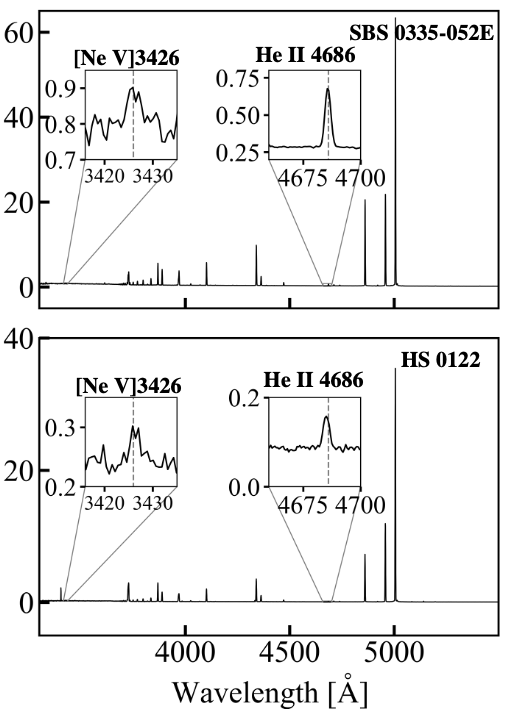} 
    \caption{Keck/LRIS spectra of SBS 0335-052E (top) and HS 0122 (bottom). 
    {The inset panels show [Ne {\sc v}]$\lambda$3426 and He{\sc ii}$\lambda$4686 emission lines for each galaxy.
    The dashed lines represent the wavelengths of the emission lines.}}
    \label{spectra_ID50_S004} 
    
\end{figure}

We use the IRAF package \citep{1993ASPC...52..173T, 1986SPIE..627..733T} for the data reduction of the LRIS spectra.
Wavelength solutions for the spectra are obtained from the HgNeArCdZnKrXe lamp, performing bias subtraction, wavelength calibration, one-dimensional (1D) spectrum extraction, flux calibration, atmospheric-absorption correction, Galactic-reddening correction, and slit loss correction.
Each spectrum is flux-calibrated with the sensitivity curve derived with Feige 34 data. 
Atmospheric absorption correction is corrected with the extinction curve at Maunakea Observatories \citep{1988BCFHT..19...16B}. 
The Galactic-reddening is corrected with the NASA/IPAC Infrared Science Archive (IRSA)\footnote{https://irsa.ipac.caltech.edu/applications/DUST/} based on the \cite{2011ApJ...737..103S} estimates. 
We produce noise frame including readout noise and photon noise of sky and object emission.  
One-dimensional spectra are extracted in the spatial width of 2 times the FWHM(H$\beta$), where FWHM(H$\beta$) is the full width half maximum value of the spatial distribution of H$\beta$ emission. 
Slit loss is estimated with the spatial profile of the H$\beta$ emission along the slit.
Figure \ref{spectra_ID50_S004} shows the one-dimensional spectra taken with LRIS.

\subsection{Flux Measurement}
\label{sec:flux_measurement}

We fit each emission line with Gaussian profiles with the $\tt{scipy.optimize}$ package, using four free parameters: The amplitude, line width, line central wavelength of Gaussian profile, and continuum of the EMPGs.
With the best-fit Gaussian profiles,
we obtain line fluxes of H$\beta$, H$\gamma$, H$\delta$, [O {\sc iii}]$\lambda \lambda$4959,5007, [O {\sc iii}]$\lambda$4363, [O {\sc ii}]$\lambda \lambda 3727,3729$, He {\sc i}$\lambda 4026$, He {\sc i}$\lambda 4471$, He {\sc ii}$\lambda$4686, [S {\sc ii}]$\lambda \lambda$6716,6731, [Ne {\sc{v}}]$\lambda$3426 and [Ne {\sc{iii}}]$\lambda$3869.
We estimate flux errors by Monte Carlo simulations with error spectra consisting of read-out noise and photon noise of sky+object emission. In our Monte Carlo simulations, we produce 1000 mock spectra, adding random errors to our galaxy spectra. We perform line flux measurements for the 1000 mock spectra as we have done for the flux measurement,
and obtain one sigma errors defined by the 68 percentile range in the distribution of the mock line fluxes.
Because we find that three pixels in the 2D spectrum of [O {\sc iii}]$\lambda$5007 emission line of SBS 0335-052E are saturated, we do not use the [O {\sc iii}]$\lambda$5007 emission line in the later analysis for SBS 0335-052E.
We calculate dust extinction $E(B-V)$, electron temperature $T_{\rm e}$, and electron density $n_{\rm e}$ iteratively in the same manner as \cite{2022ApJ...925..111I} with H$\beta$, H$\gamma$ and H$\delta$. 
We derive three intrinsic Balmer line raitos of H$\beta$, H$\gamma$, H$\delta$ with PyNeb v.1.1.15\citep{2015A&A...573A..42L} and calculate three $E(B-V)$ values, assuming the case B recombination and the dust attenuation curve of \cite{2000ApJ...533..682C}. 
An error of $E(B-V)$ is derived based on a $\chi^2$ value calculated with $\tt{least\_squares}$.
The dust-corrected fluxes of SBS 0335-052E and HS 0122+0743 are summarized in Table \ref{flux_obs}.  

{We point out that the $\lbrack$Ne\ {\sc{v}}]$\lambda$3426/[Ne\ {\sc{iii}}]$\lambda$3869 of SBS 0335-052E is consistent with those reported in \cite{Izotov2009} and \cite{2005ApJS..161..240T}, whose observations are conducted in September 2003 and February 2004, respectively.}

{He {\sc ii}$\lambda$4686/H$\beta$ emission line ratios in the dwarf galaxies are relatively lower than typical AGNs. 
The reason of the low He {\sc ii}/H$\beta$ emission is explained by the fact that, in dwarf galaxies, stellar radiation is much stronger than AGN radiation at the energy level of the hydrogen ionizing photons, $\sim$13.6 eV, producing H$\beta$ emission. 
It is claimed that non-stellar radiation such as AGNs are thought to contribute only $\sim$10 per cent to the total luminosity of ionizing radiation of dwarf galaxies with the [Ne {\sc v}]$\lambda$3426 line detection (e.g. \citealt{2012MNRAS.427.1229I,2021MNRAS.508.2556I}). }

\begin{deluxetable*}{lccccccccc}
\tablecaption{Sample Properties}
\tablewidth{700pt}
\tabletypesize{\scriptsize} 
\label{table:sample_properties}
\tablehead{
Name of the Galaxy & $z$  & $i$ & log($M*$)  & $12 + \log ({\rm O/H})$ & $F({\rm H}\beta)$ & $W1$ & $W1-W2$ & $W2-W3$ & {\rm References} \\
& &(mag)& ($M_\odot$) & & ($10^{-16} {\rm erg \ s^{-1} \ cm^{-2}}$)&Mag & Mag & Mag &\\
& (1)&(2)& (3)& (4)& (5) & (6) & (7)& (8) & (9)
}

\startdata
SBS 0335--052E      & 0.01352 & 18.92    & 7.7                     & 7.29$\pm$0.02          & 690.0$\pm$0.4$\ddag$         & 14.52    & 1.99     & 4.85     & (a), (b), (c), (d), (e)\\
HS 0122+0743        & 0.0097  & 19.31    & 6.39$\pm$0.23           &  7.60                  & 262.7$\pm$0.4$\ddag$         & 15.48    & 0.69     & 4.54     & (f), (g), (e)\\
J104457             & 0.013   & 18.73    & 6.80                    & 7.45                   & 95$\pm$1.0                   & 16.07    & 1.00     & 4.88     & (h), (g)\\
J1222+3602          & 0.3011  & 21.47    & 8.4$\pm$0.4             & 7.686$\pm$0.024        & 10.5$\pm$0.1                 & 16.25    & 1.20     & 3.67     & (i), (g), (j)  \\
W1702+18            & 0.0425  & 17.98    & 8.4$\dag$               & 7.753$\pm$0.020        & 196.0$\pm$0.2                & 14.29    & 2.25     & 4.44     & (i), (g)\\
Tol 1214-277        & 0.0260  & 19.52    & 7.5$\dag$               &  7.56$\pm$0.01         & 189$\pm$1                    & 16.72    & 0.88     & 5.03     & (a), (k)\\
J1205+4551          & 0.0654  & 19.62    & 6.84                    & 7.460$\pm$0.022        & 44.4$\pm$0.1                 & 15.14    & 1.59     & 3.64     & (i), (g), (l)\\
J0344-0106          & 0.2707  & 21.73    & 8.6$\dag$               & 7.688$\pm$ 0.028       & 5.86$\pm$ 0.2                & 17.36    & 0.88     & 3.94     & (i), (g)\\
J024009.10+010334.5 & 0.1956  & 19.64    & 9.0$\dag$               & $\cdots$               & 4.1$\pm$0.3$\ddag$           & 16.24    & 1.13     & 2.98     & (m), (g), (e)\\
ID 6355             & 7.665   & $\cdots$ & $8.77^{+0.08}_{-0.01}$  & $8.35^{+0.11}_{-0.08}$ & $(5.33\pm0.12)\times10^{-2}$ & $\cdots$ & $\cdots$ & $\cdots$ & (n), (o)\\
\hline
\enddata
\tablecomments{
(1): Redshift. (2): $i$-band magnitude. The magnitudes are the iMeanPSF magnitudes of Pan-STARRS1 DR2 (\citealt{2016arXiv161205560C, 2020ApJS..251....7F}) for Tol 1214-277, I Cousins band magnitude obtained by Cerro Tololo Inter-American Observatory (CTIO) for SBS 0335-052E, and cModel magniutudes of SDSS DR15 (\citealt{2019ApJS..240...23A}) for the others. (3): Stellar mass. (4) Gas-phase metallicity derived by the direct $T_{\rm{e}}$ method. (5): H$\beta$ flux. {(6): WISE $W1$ band magnitudes in Vega magnitudes.} {(7): WISE $W1-W2$ colors.} {(8): WISE $W2-W3$ colors.} (9): References for values in column (1)-({8}): (a) NASA/IPAC Extragalactic Database (NED), (b) \cite{2005ApJ...621..104T}, (c) \cite{Pustilnik2004}, (d) \cite{Izotov2009}, (e) This work, (f) \cite{Filho2013}, (g) \cite{2019ApJS..240...23A}, (h) \cite{2021ApJ...922..170B}, (i) \cite{2021MNRAS.508.2556I}, (j) \cite{2021MNRAS.504..543B}, (k) \cite{Izotov2004}, 
(l) \cite{2017MNRAS.471..548I}, (m) \cite{2011ApJ...739...28X}, (n) \cite{2023arXiv230112825N}, and (o) Isobe et al. in prep.
($\dag$) The stellar masses calculated with the $i$-band magnitude and the relation of stellar masses and $i$-band absolute magnitude given by \cite{2021ApJ...918...54I}.
($\ddag$) The H$\beta$ fluxes obtained by this work. {The WISE magnitudes are taken from NED.}
}
\end{deluxetable*}

\begin{longrotatetable}
\begin{deluxetable*}{lccccccccc}
\tablecaption{Fluxes and Flux Ratios \label{chartable}}
\tablewidth{700pt}
\tablehead{ion&SBS 0335-052E & HS 0122+0743 &J104457 &J1222+3602 & W1702+18 & Tol 1214-277 & J1205+4551 & J0344-0106 & J024009.10+010334.5}
\tabletypesize{\scriptsize}
\label{flux_obs}
\startdata
\multicolumn{10}{c}{$F_{\lambda}/F_{{\rm H}\beta} \times 100 $}\\
\hline
$\lbrack$O\ {\sc{ii}}]$\lambda\lambda$3727,3729&24.9$\pm$1.4&62.8$\pm$2.0&26.37$\pm$0.27&36.42$\pm$1.21&63.04$\pm$2&28$\pm$0.6&18.26$\pm$0.59&20.17$\pm$0.74&348$\pm$128\\
He\ {\sc{i}}$\lambda$4026&1.66$\pm$0.07&1.68$\pm$0.07&1.72$\pm$0.03&2.46$\pm$0.2&1.81$\pm$0.07&1.5$\pm$0.3&1.97$\pm$0.08&1.87$\pm$0.23&$\cdots$\\
H$\delta$&27.0$\pm$0.9&26.1$\pm$0.6&25.59$\pm$0.42&26.28$\pm$0.89&28.34$\pm$0.86&26.6$\pm$0.5&27.5$\pm$0.84&25.47$\pm$0.89&29.9$\pm$9.0\\
H$\gamma$&48.7$\pm$1.1&48.1$\pm$0.7&46.55$\pm$0.67&48.25$\pm$1.49&48.64$\pm$1.43&47.4$\pm$0.8&46.99$\pm$1.44&45.93$\pm$1.46&62.2$\pm$12.5\\
$\lbrack$O\ {\sc{iii}}]$\lambda$4363&10.9$\pm$0.2&13.4$\pm$0.2&13.51$\pm$0.21&29.52$\pm$0.91&15.41$\pm$0.45&16.6$\pm$0.3&12.87$\pm$0.38&16.36$\pm$0.63&53.1$\pm$10.2\\
He\ {\sc{i}}$\lambda$4471&3.77$\pm$0.07&3.84$\pm$0.07&3.97$\pm$0.09&4.68$\pm$0.21&3.97$\pm$0.12&3$\pm$0.2&4.02$\pm$0.13&$\cdots$&$\cdots$\\
He\ {\sc{ii}}$\lambda$4686&2.22$\pm$0.02&1.13$\pm$0.04&1.8$\pm$0.04&1.13$\pm$0.11&1.96$\pm$0.07&5$\pm$0.1&3.24$\pm$0.11&3.41$\pm$0.16&24.3$\pm$8.8\\
H$\beta$&100.00$\pm$0.06&100.00$\pm$0.15&100.0$\pm$1.4&100.00$\pm$2.92&100.00$\pm$2.86&100.0$\pm$1.5&100.00$\pm$2.86&100.00$\pm$2.91&100.0$\pm$6.2\\
$\lbrack$O\ {\sc{iii}}]$\lambda$4959&104.6$\pm$0.4&161.4$\pm$0.4&143$\pm$1.5&281.07$\pm$8.16&205.07$\pm$5.87&173.4$\pm$2.6&139.15$\pm$3.98&199.09$\pm$5.76&105.9$\pm$5.0\\
$\lbrack$O\ {\sc{iii}}]$\lambda$5007&$\cdots$&479.7$\pm$1.7&427.5$\pm$4.3&832.21$\pm$24.2&609.85$\pm$17&521.9$\pm$7.6&414.63$\pm$11.9&608.79$\pm$17.6&305.7$\pm$12.8\\
\hline
\multicolumn{10}{c}{Other Line Ratios}
\\ \hline
$\lbrack$S\ {\sc{ii}}]$\lambda$6716/$\lambda$6731&1.24$\pm$0.1&1.32$\pm$0.07&1.23$\pm$0.04&0.996$\pm$0.081&1.16$\pm$0.052&1.15$\pm$0.12&1.02$\pm$0.055&1.39$\pm$0.13&1.22$\pm$0.37\\
$\lbrack$Ne\ {\sc{v}}]$\lambda$3426/[Ne\ \sc{iii}]$\lambda$3869&0.0286$\pm$0.0049&0.0189$\pm$0.0052&0.00347$\pm$0.0032&0.0212$\pm$0.0028&0.00889$\pm$0.0024&0.0844$\pm$0.013&0.0297$\pm$0.0095&0.0394$\pm$0.0079&0.786$\pm$0.32\\
\enddata
\tablecomments{Extinction-corrected emission-line fluxes and emission-line ratios. The fluxes are normalized with H$\beta$ = 100. } 
\end{deluxetable*}
\end{longrotatetable}

\section{Sample}

\label{sec:sample}

\subsection{Making Our Sample}

We use eight dwarf galaxies with detection of [Ne {\sc v}]$\lambda$3426 emission lines, two of which, SBS 0335-052E and HS 0122, are 
taken from our observations,
while six dwarf galaxies are taken from the literature \citep{2021MNRAS.508.2556I, Izotov2004, 2021ApJ...922..170B}.
%
{For testing our mass estimate technique, we search for low-mass SMBHs and IMBHs with a [Ne {\sc v}]$\lambda$3426 emission line detection in the database of SDSS spectra. 
Because we have accretion disk model SEDs with masses only up to $M_{\rm BH}\lesssim 10^{6.5} \ M_\odot$, we cannot use SMBHs with $M_{\rm BH}\gtrsim 10^{6.5} \ M_\odot$. 
We find an IMBH dubbed J024009.10+010334.5 (hereafter J024009; \citealt{2011ApJ...739...28X}), harboring an AGN with a low BH mass of $M_{\rm BH}=10^{5.75}\ M_\odot$ that shows a [Ne {\sc{v}}]$\lambda$3426-line detection.
We also find some candidate galaxies in the LBT archive. However, we do not have access to the spectra.  In summary, we find only one IMBH, J024009, for the testing purposes.}
The emission line fluxes of J024009 are measured in the same method explained in Section \ref{sec:flux_measurement}. 
The extinction correction is applied to J024009 with H$\beta$/H$\gamma$ ratio, assuming electron temperature of $T_{\rm e} = 8,000$ K and deriving electron density $n_{\rm e}$ with [S {\sc ii}] ratio.

We add a star forming galaxy at $z=7.665$, ID 6355,  recently identified in James Webb Space Telescope (JWST) observations with [Ne {\sc iv}]$\lambda$2424 line detection to our sample (\citealt{2023MNRAS.518..425C, 2022arXiv220807467B}).
The data was taken by JWST/NIRSpec in Early Release Observations (ERO; \citealt{2022ApJ...936L..14P}).
The spectra were taken with medium resolution gratings/filters of G235M/F170LP and G395M/F290LP, which cover the wavelength range of 1.7--3.1 and 2.9 -- 5.1 ${\rm \mu}$m, respectively. 
We use the data reduced in \cite{2023arXiv230112825N}. 
Emission line fluxes are obtained by the same manner as Isobe et al. in prep. by fitting the Gaussian distribution convolved with the line-spread function defined in \cite{2023arXiv230106811I}. 
The redshift and velocity dispersion are fixed to the values obtained from [O {\sc iii}]$\lambda$5007 line.
We calculate the dust extinction $E(B-V)$ with H$\beta$, H$\gamma$, and H$\delta$ line fluxes, with $T_{\rm e}$ and $n_{\rm e}$ taken from \cite{2023arXiv230112825N} and \cite{2023arXiv230106811I}, respectively. We obtain the value of $E(B-V) = 0.060\pm 0.16$
and conduct extinction correction with this value. 
The obtained emission line fluxes and ratios are summarized in Table \ref{table:flux_JWST}. 

To summarize, our sample is composed of a total of 10  galaxies: 8 local star-forming galaxies, 1 known local low-mass AGN, and 1 high-$z$ star forming galaxy. Properties of the 10 galaxies are summarized in Table \ref{table:sample_properties}. 

\begin{figure*}
    \centering
    \includegraphics[width=15 cm]{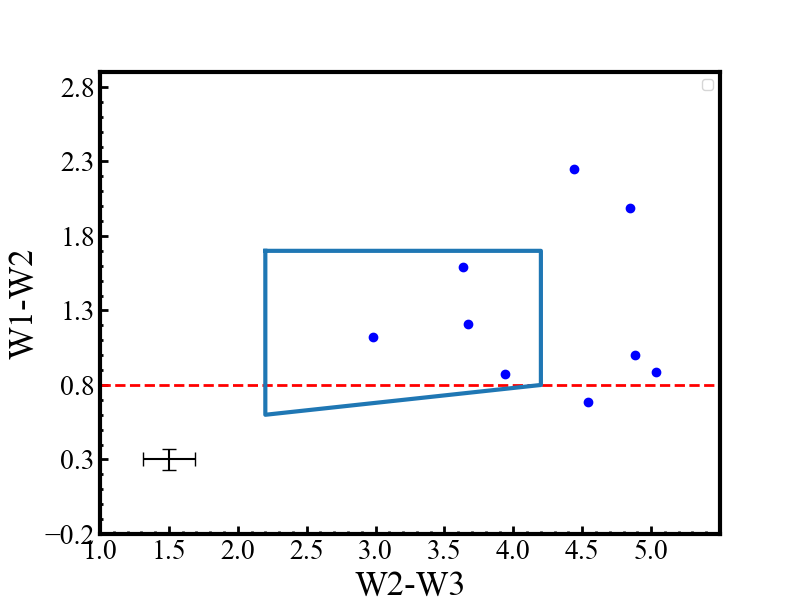}
    \caption{{WISE color-color diagram for the 9 dwarf galaxies (blue circles) in our sample. Red dashed line (blue box) represent an AGN criterion of \cite{2012ApJ...753...30S} (\citealt{2011ApJ...735..112J}). 
    We plot the uncertainties with the error bars shown in \cite{2016ApJ...832..119H} in the bottom-left corner.}} 
    \label{fig:WISE}
\end{figure*}

{We investigate the mid-infrared (MIR) colors of the dwarf galaxies, because MIR colors are good indicators of the presence of hot-dust, which could originate from AGN activities.
We use $W1 \ (3.4 \ {\rm \mu m})$, $W2 \ (4.6 \ {\rm \mu m})$, and $W3 \ (12\ {\rm \mu m})$ photometry in the AllWISE catalogue (\citealt{2010AJ....140.1868W}). 
We show the mid-infrared (MIR) colors of the dwarf galaxies and AGN criteria of \cite{2012ApJ...753...30S} and \cite{2011ApJ...735..112J} in Figure \ref{fig:WISE}.}
{We point out that the MIR colors of the dwarf galaxies, except for HS 0122, satisfy an AGN criterion ($W1-W2>0.8$) proposed in \cite{2012ApJ...753...30S} (cf. \citealt{2016ApJ...832..119H}). 
Moreover, four of the dwarf galaxies, J1205, J0344, J1222, and J024009 satisfy an AGN criterion proposed in \cite{2011ApJ...735..112J}.}

{
Some of the dwarf galaxies show temporal variability. 
J1222 is claimed to harbor an AGN on the basis of a presence of H$\alpha$ broad line lasting for more than 10 years \citep{2021MNRAS.504..543B,2021MNRAS.508.2556I}.
J1205 show temporal variability in $W1$ and $W2$ photometry in NEOWISE data \citep{2014ApJ...792...30M, 2023ApJ...945..157H}. 
SBS 0335-052E show a temporal variability in NEOWISE data \citep{2023arXiv230403726H}. 
}

\section{Method}
\label{sec:modeling}

As explained in Section 1, our model follow the method of \citetalias{2022ApJ...930...37U} with an improvement of stellar radiation by using BPASS v.2.2.1 single models \citep{2018MNRAS.479...75S}.
\subsection{Photoionization Models}
\label{sec:cloudy_models}
We used version 17.02 of CLOUDY, last described by \cite{2017RMxAA..53..385F}, to calculate the model emission line fluxes.
We stop the radiative transfer calculation when a neutral hydrogen column density $N_{\rm{H \ I}}$ reaches $N_{\rm{H \ I}} = 10^{21} \ {\rm{cm}^{-2}}$. We normalize the output emission line fluxes by model $\rm{H 
\beta}$ flux, for convenience of calculation. 
The detailed settings of CLOUDY models are explained below.

\subsubsection{Geometry and Density}
We assume a photionization model having closed and spherical geometry, with a constant hydrogen density. 
We adopt 
the hydrogen density $n_{\rm{H}}$ ranging in $10^{0.5}-- 10^5 \ \rm{cm}^{-3}$. 
The inner and outer radii of the gas cloud $R_{\rm in}$ and $R_{\rm out}$ are fixed at the default values.
\subsubsection{Ionizing Spectra}
\label{sec:ionizing_spectra}
We use a combination of stellar and power-law radiation for an input ionizing spectrum. 
The input ionizing spectrum is defined by 
\begin{equation}
\label{eq:defin_spectra}
    F_\nu = S(\nu, t) + C_{\rm{mix}} P(\nu, \alpha_{\rm{X}}),
\end{equation}
where $S(\nu,t)$ is the stellar spectrum of a BPASS single-burst model at a stellar age $t$ and frequency $\nu$, $C_{\rm{mix}}$ is the mixing parameter, and $P(\nu, \alpha_{\rm{X}})$ is the power-law spectrum. {We use BPASS single-burst models because \citetalias{2022ApJ...930...37U} have pointed out that BPASS binary models cannot reproduce ionizing spectra of some EMPGs.}
Here, $P(\nu, \alpha_{\rm{X}})$ is expressed as
 \begin{equation}
 \label{eq:power-law}
     P(\nu, \alpha_{\rm X}) = {\nu}^{\alpha_{\rm{x}}} {\rm{exp}}(-h\nu/E_{\rm{hc}}) {\rm{exp}}(-E_{\rm{lc}}/h\nu), 
 \end{equation}
where $\alpha_{\rm X}$ and $h$ are the power-law index and the Planck constant, respectively. 
To avoid strong free-free (pair-creation and Compton) heating, we set the lower (higher) energy cut $E_{\rm{lc}}$ ($E_{\rm{hc}}$) for the power-law spectrum. Here, we set $E_{\rm{lc}} = 0.1$ Ryd and $E_{\rm{hc}} = 10,000$ Ryd.
We use normalized mixing parameter $a_{\rm{mix}}$ as an input parameter of CLOUDY, 
where $a_{\rm{mix}}$ is defined as
\begin{equation}
\label{eq:C_mix}
    a_{\rm{mix}} \equiv C_{\rm{mix}} \frac{P(\nu = 1 \ {\rm{Ryd}}/h, \alpha_{\rm{X}})}{S(\nu = 1 \ {\rm{Ryd}}/h, t)}.
\end{equation}
We adopt 
the parameter ranges of
$6 \leq \rm{log}\ (t_{\rm stellar}/{\rm yr}) \leq 8$, $-3 \leq \alpha_{\rm{X}} \leq 1$, and $-4 \leq {\rm{log}}\ a_{\rm{mix}} \leq 3$, respectively.
\subsubsection{Chemical Abundance}
In our model, we set chemical abundances with oxygen abundance (hereafter referred to as $Z$).
We scale all chemical abundances linearly with solar abundance ratios given by \cite{2010Ap&SS.328..179G}, with the exception of helium, carbon, and nitrogen. 
We calculate helium and carbon (nigrogen) abundances with nonlinear formula given by 
\cite{2006ApJS..167..177D} \citep{2012MNRAS.426.2630L}. 
We allow chemical abundances to vary in the range of $-3 \leq {\rm{log}}(Z/Z_\odot) \leq 0$.

\subsubsection{Ionization Parameter}
We define the ionization parameter $U$ by
\begin{equation}
    U \equiv \frac{Q(\rm{H}^0)}{4\pi R_{\rm{S}}^2 n_{\rm{H}} c},
\end{equation}
where $Q(\rm{H}^0)$ is the intensity of hydrogen ionizing photons
$R_{\rm{S}}$ is the Str$\ddot{\rm{o}}$mgren radius,
and $c$ is the speed of light.
We substitute $R_{\rm{in}}$ for $R_{\rm{S}}$ to calculate the ionizing parameter $U$. 
We limit the ionizing parameter in the range of $-5 \leq {\rm{log}}\  U \leq -0.5$.

\subsection{MCMC Parameter Estimates}
\label{sec:param_estimate}
To estimate the best-fit parameters of our CLOUDY photoionization model, we use $\tt{emcee}$, a Python implementation of an affine invariant MCMC sampling algorithm \citep{2013PASP..125..306F}. 
We maximize the log-likelihood function given by
\begin{equation}
\label{eq:lnL}
    \ln \mathcal{L} = - \frac{1}{2} \sum_{\lambda \in \Lambda} \left[ \left( \frac{F_{\lambda, {\rm{obs}}} - F_{\lambda, {\rm{mod}}}}{\sigma_{\lambda}}\right)^2 + {\rm{ln}} (2\pi \sigma_{\lambda}^2)\right],
\end{equation}
where $\Lambda$ is the set of emission lines, $F_{\lambda, {\rm{obs}}}$ ($F_{\lambda, {\rm{mod}}}$) are the observed (model) emission line fluxes at wavelength $\lambda$, and $\sigma_{\lambda}$ are the observed errors of the emission line fluxes at $\lambda$ normalized at $F_{\rm H\beta, obs} = 100$. 
Here, $F_{\lambda, {\rm{mod}}}$ is defined by 
\begin{equation}
    F_{\lambda, {\rm{mod}}} = N_{{\rm{H}}\beta} \frac{F_{\lambda, {\rm{cloudy}}}}{F_{{\rm{H}}\beta, {\rm{cloudy}}}}
\end{equation}
where $N_{{\rm{H}}\beta}$ is the normalization factor for an H$\beta$ emission line. 
$F_{\lambda, {\rm{cloudy}}}$ is the output value of CLOUDY for the emission line flux at $\lambda$. 
To account for the error of H$\beta$ fluxes, we add $N_{{\rm{H}}\beta}$ as a free parameter by letting it vary in {[100 - 3$\sigma_{\rm H\beta}$, 100 + 3$\sigma_{\rm H\beta}$]}. 

For the prior distribution, we use uniform distribution. 
We summarize the prior distributions of all 7 free parameters in Table \ref{prior}. 
We set 40 walkers and run the MCMC sampling algorithm for $\sim$1000 steps, sampling $\sim$40,000 parameter sets in total.
We define the ``best-fit'' parameter set as a parameter set having maximum likelihood value among all the sampled parameter sets.
The uncertainty of the parameters is defined by the range of the parameter sets satisfying a condition  
\begin{equation}
\label{eq:lnL_3sigma}
    {\rm ln }\mathcal{L} \geq {\rm ln}\mathcal{L}_{3\sigma} = - \frac{1}{2} \sum_{\lambda \in \Lambda} \left[ \left(\frac{3\sigma_{\lambda}}{\sigma_{\lambda}} \right)^2 + {\rm ln}(2\pi \sigma_{\lambda}^2)\right].
\end{equation}
We note that the uncertainty provided in Equation \ref{eq:lnL_3sigma} is only a rough standard, and the actual uncertainty can be more rigorously defined.
\section{Results}
\label{sec:result}

\begin{deluxetable}{lc}
\tablecaption{Fluxes and Flux Ratios \label{JWST_fluxes}}
\tablewidth{700pt}
\tablehead{ion& ID 6355}
\tabletypesize{\scriptsize}
\label{table:flux_JWST}
\startdata
\multicolumn{2}{c}{$F_{\lambda}/F_{{\rm H}\beta} \times 100 $}\\
\hline 
$\lbrack$O\ {\sc{ii}}]$\lambda\lambda$3727&56.4$\pm$10.9\\
$\lbrack$O\ {\sc{ii}}]$\lambda\lambda$3729&52.3$\pm$10.4\\
He\ {\sc{i}}$\lambda$4026&5.4$\pm$1.8\\
H$\delta$&26.1$\pm$3.5\\
H$\gamma$&47.6$\pm$4.1\\
$\lbrack$O\ {\sc{iii}}]$\lambda$4363&8.7$\pm$2.0\\
He\ {\sc{i}}$\lambda$4471&5.1$\pm$1.8\\
He\ {\sc{ii}}$\lambda$4686& $<2.9$\\
H$\beta$&100.0$\pm$2.3\\
$\lbrack$O\ {\sc{iii}}]$\lambda$4959&256.8$\pm$8.3\\
$\lbrack$O\ {\sc{iii}}]$\lambda$5007&783.4$\pm$15.1\\
\hline
\multicolumn{2}{c}{Other Line Ratios}
\\ \hline
$\lbrack$Ne\ {\sc{iv}}]$\lambda\lambda$2422,2424/[Ne\ \sc{iii}]$\lambda$3869&0.370$\pm$0.155\\
\enddata
\tablecomments{Extinction-corrected emission-line fluxes and emission-line ratios of ID 6355. The fluxes are normalized with H$\beta$ = 100. We show the 3$\sigma$ upper limit flux for the He\ {\sc{ii}}$\lambda$4686 line of ID 6355.} 
\end{deluxetable}

\subsection{Best-Fit Parameters for the Nebular and Ionizing Spectral Models}
\label{sec:best-fit}
  We use observed $\sim$14 emission lines originating from hydrogen, helium, oxygen, sulfur, and neon ions summarised in
  Table \ref{flux_obs} for the 9 local galaxies. 
  We do not include H$\alpha$ fluxes in our analysis 
  because there may remain
  flux calibration systematics between 
  our LRIS blue and red channel data, 
  in the latter of which only H$\alpha$ emission falls.
  To remove uncertainties of abundance ratio differences,
  we use the ratios, [S {\sc{ii}}]$\lambda$6716/[S {\sc{ii}}]$\lambda$6731 and [Ne {\sc{v}}]$\lambda$3426/[Ne {\sc{iii}}]$\lambda$3869, for sulfur and neon lines, respectively.
  For ID 6355, we use emission line fluxes and flux ratios listed in Table \ref{JWST_fluxes}. Because [Ne {\sc v}]$\lambda$3426 line is not detected in ID 6355, we use [Ne {\sc iv}]$\lambda$2424/[Ne {\sc iii}]$\lambda$3869 instead of [Ne {\sc{v}}]$\lambda$3426/[Ne {\sc{iii}}]$\lambda$3869. Because [He {\sc ii}]$\lambda$4686 emission line is not detected in ID 6355, we search the best-fit parameters that are consistent with three sigma non-detection of the [He {\sc ii}]$\lambda$4686 line.

\begin{figure*}
    \centering
    \includegraphics[width=15 cm]{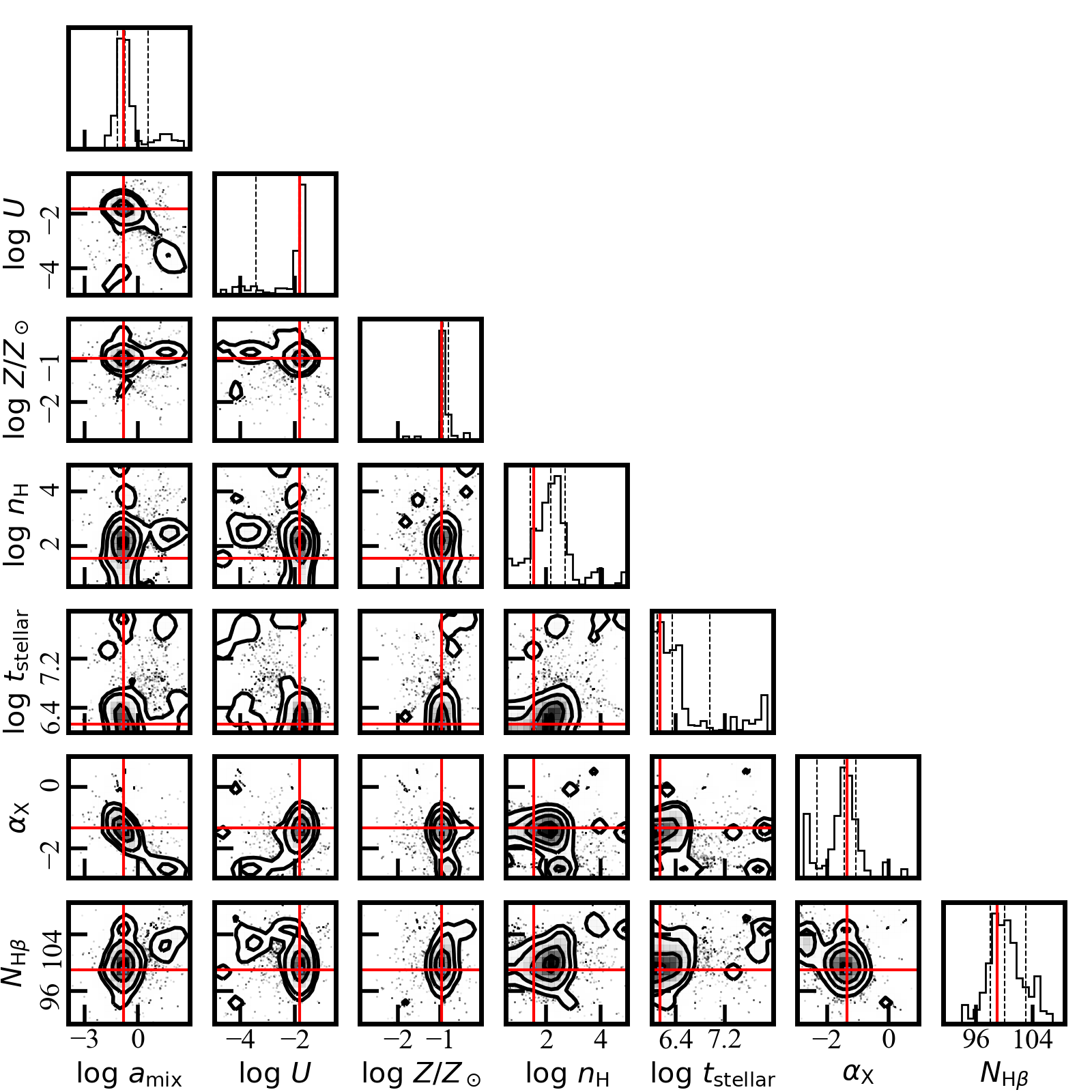}
    \caption{Posterior PDF of the model parameters for J0344. Two-dimensional (one-dimensional) probability distribution for each parameter is shown on the off-diagonals (along the diagonals). The darker regions on the joint probability distributions indicate the higher density of the sampled parameter sets. The red solid lines (black dashed lines) represent the best-fit values (68$\%$ confidence range) of model parameters.}
    \label{corner_ID50}
\end{figure*}

  We compare the observed emission line fluxes and ratios with the photoionization models (Section \ref{sec:cloudy_models}), performing the MCMC parameter estimates (Section \ref{sec:param_estimate}).
  In Figure \ref{corner_ID50}, we present 
  the posterior probability distribution function (PDF) 
  for the J0344 galaxy as an example, and determine the best-fit parameters
  and the associated errors.
  We also obtain the best-fit parameters and the uncertainties using Equation \ref{eq:lnL_3sigma} for all of our dwarf galaxies that are summarized in Table \ref{best-fit_value}. 

 {
 The range of hydrogen density used in the MCMC technique does not include the critical density of the [Ne {\sc v}]$\lambda$3426 emission line ($n_{\rm H} \sim n_{\rm e} = 1.5\times 10^7$ cm$^{-3}$), and the best-fit parameters could be found outside the current parameter range.
 We extend the range of hydrogen density up to ${\rm log} (n_{\rm H} {\rm /cm^{-3}}) = 10$, which includes the critical density of the [Ne {\sc v}]$\lambda$3426 emission line. We also extend the ionization parameter up to  ${\rm log} U$ =1,  encompassing the range of ionization parameters typically observed in the circum-nuclear environment of AGNs ( ${\rm log} U \sim (-3) - 1$; \citealt{1990agn..conf...57N}). 
 We conduct MCMC calculations for Tol 1214 that has reliably high S/N ratios for high ionization lines including [Ne {\sc v}]$\lambda$3426 with the new parameter ranges, and obtain the best-fit parameters: ${\rm log} U = -1.93$ and ${\rm log} n_{\rm H} {\rm /cm^{-3}} = 2.54$. These parameters do not change the conclusions of the $M_{\rm BH}$ value.
 }

\begin{table}
\begin{center}
  \caption{Prior Distributions of Free Parameters}
  \label{table:data_type}
  \centering
  \begin{tabular}{lcc}
    \hline \\ \hline
      & Parameter  & Prior Range \\
    \hline
    (1)  & $ {\rm log} a_{\rm mix}$& [--4,3] \\
    (2)  & $ {\rm log} t_{\rm stellar}{{\rm /yr}}$  & [6,8]\\
    (3)  & $\alpha_{\rm X}$ &  [--3,1]\\
    (4)  & ${\rm log} U$  & [--5,--0.5] \\
    (5)  & ${\rm log} n_{\rm H} {\rm /cm^{-3}}$ & [0.5,5]\ \\ 
    (6)  & ${\rm log} Z/Z_{\odot}$ & [--3,0]\\
    (7)  & $N_{{\rm{H}}\beta}$ & [100 - 3$\sigma_{\rm H\beta}$,100 + 3$\sigma_{\rm H\beta}]$ \\
    \hline
    \label{prior}
\end{tabular}
  \end{center}
\footnotesize{{\sc Note}--- (1): Ratio of stellar flux to the power-law flux at 1 Ryd. (2): Stellar age. (3): Power-law index. (4): Ionization paramter. (5): Hydrogen density. (6): Gas-phase metallicity. (7): Normalized factor for an H$\beta$ emission line.} 
\end{table}

\begin{table*}[hbtp]
  \begin{center}

  \caption{Best-fit Parameters}
  \label{best-fit_value}
  \centering
  \begin{tabular}{lccccccc}
\hline \\ \hline
Parameters & $ {\rm log} a_{\rm mix}$& $ {\rm log} t_{\rm stellar}{{\rm /yr}}$ & $\alpha_{\rm X}$  & ${\rm log} U$ &  ${\rm log} n_{\rm H} {\rm /cm^{-3}}$ & ${\rm log} Z/Z_{\odot}$ & $N_{{\rm{H}}\beta}$\\
\hline
\multicolumn{8}{c}{Best-fit Parameters}\\
\hline
SBS 0335-052E    &$-2.76^{+0.75}_{-0.083}$&$6.63^{+0.11}_{-0.3}$&$0.29^{+0.078}_{-0.78}$&$-2.09^{+0.22}_{-0.14}$&$3.37^{+0.15}_{-0.45}$&$-1.27^{+0.028}_{-0.04}$&$100.0^{+0.16}_{-0.17}$\\
HS 0122+0743     &$-3.4^{+0.26}_{-0.099}$&$6.67^{+0.051}_{-0.042}$&$0.487^{+0.085}_{-0.22}$&$-2.01^{+0.23}_{-0.12}$&$3.25^{+0.29}_{-0.28}$&$-0.988^{+0.028}_{-0.027}$&$100.0^{+0.38}_{-0.44}$\\
J104457          &$-2.63^{+0.38}_{-0.25}$&$6.09^{+0.23}_{-0.092}$&$0.214^{+0.17}_{-0.35}$&$-2.1^{+0.07}_{-0.033}$&$2.91^{+0.27}_{-0.31}$&$-1.13^{+0.029}_{-0.026}$&$102.0^{+1.8}_{-3.6}$\\
J1222+3602       &$-3.7^{+0.55}_{-0.3}$&$6.41^{+0.28}_{-0.38}$&$0.683^{+0.094}_{-0.3}$&$-1.92^{+0.25}_{-0.17}$&$3.48^{+0.76}_{-0.75}$&$-0.896^{+0.16}_{-0.071}$&$105.0^{+3.5}_{-6.7}$\\
W1702+18         &$-2.68^{+0.68}_{-0.4}$&$6.09^{+0.49}_{-0.087}$&$0.266^{+0.21}_{-0.79}$&$-2.22^{+0.18}_{-0.11}$&$2.86^{+0.44}_{-1.3}$&$-0.879^{+0.17}_{-0.056}$&$106.0^{+2.4}_{-12.0}$\\
Tol 1214-277     &$-1.48^{+0.88}_{-0.43}$&$6.62^{+0.11}_{-0.59}$&$-0.615^{+0.47}_{-0.75}$&$-1.95^{+0.12}_{-0.21}$&$2.57^{+0.98}_{-1.7}$&$-1.04^{+0.082}_{-0.048}$&$101.0^{+3.0}_{-2.0}$\\
J1205+4551       &$-1.66^{+4.6}_{-0.72}$&$6.3^{+1.0}_{-0.3}$&$-0.554^{+0.65}_{-2.2}$&$-2.0^{+0.31}_{-2.7}$&$3.2^{+0.6}_{-0.68}$&$-1.18^{+0.2}_{-0.051}$&$104.0^{+4.1}_{-8.0}$\\
J0344-0106       &$-0.801^{+1.8}_{-1.1}$&$6.13^{+0.49}_{-0.13}$&$-1.34^{+0.9}_{-1.2}$&$-1.82^{+0.13}_{-0.8}$&$1.53^{+1.7}_{-1.0}$&$-0.962^{+0.24}_{-0.082}$&$99.0^{+6.7}_{-4.9}$\\
J024009.10+010334009&$0.542^{+2.5}_{-4.5}$&$7.78^{+0.22}_{-1.8}$&$-0.202^{+0.52}_{-2.2}$&$-2.49^{+2.0}_{-2.5}$&$2.3^{+2.7}_{-1.8}$&$-1.63^{+1.6}_{-0.14}$&$103.0^{+15.0}_{-20.0}$\\
JWST ID 6355     &$-1.62^{+1.8}_{-2.4}$&$6.27^{+0.54}_{-0.27}$&$-0.626^{+1.3}_{-2.4}$&$-2.2^{+1.6}_{-0.42}$&$3.24^{+1.7}_{-2.7}$&$-0.467^{+0.46}_{-0.36}$&$100.0^{+5.8}_{-6.2}$\\
\hline
\\ \hline
\end{tabular}
  \end{center}
\footnotesize{{\sc Note}--- The best-fit parameters are the parameter sets that maximize the $\ln \mathcal{L}$ values in equation (\ref{eq:lnL}). The uncertainties indicate the parameter sets satisfying the condition shown in Equation (\ref{eq:lnL_3sigma}). (1): Ratio of stellar flux to the power-law flux at 1 Ryd. (2): Stellar age. (3): Power-law index. (4): Ionization paramter. (5): Hydrogen density. (6): Gas-phase metallicity. (7): Normalized factor for an H$\beta$ emission line.}

\end{table*}

\begin{figure*}
    \centering
    \includegraphics[width = 15cm]{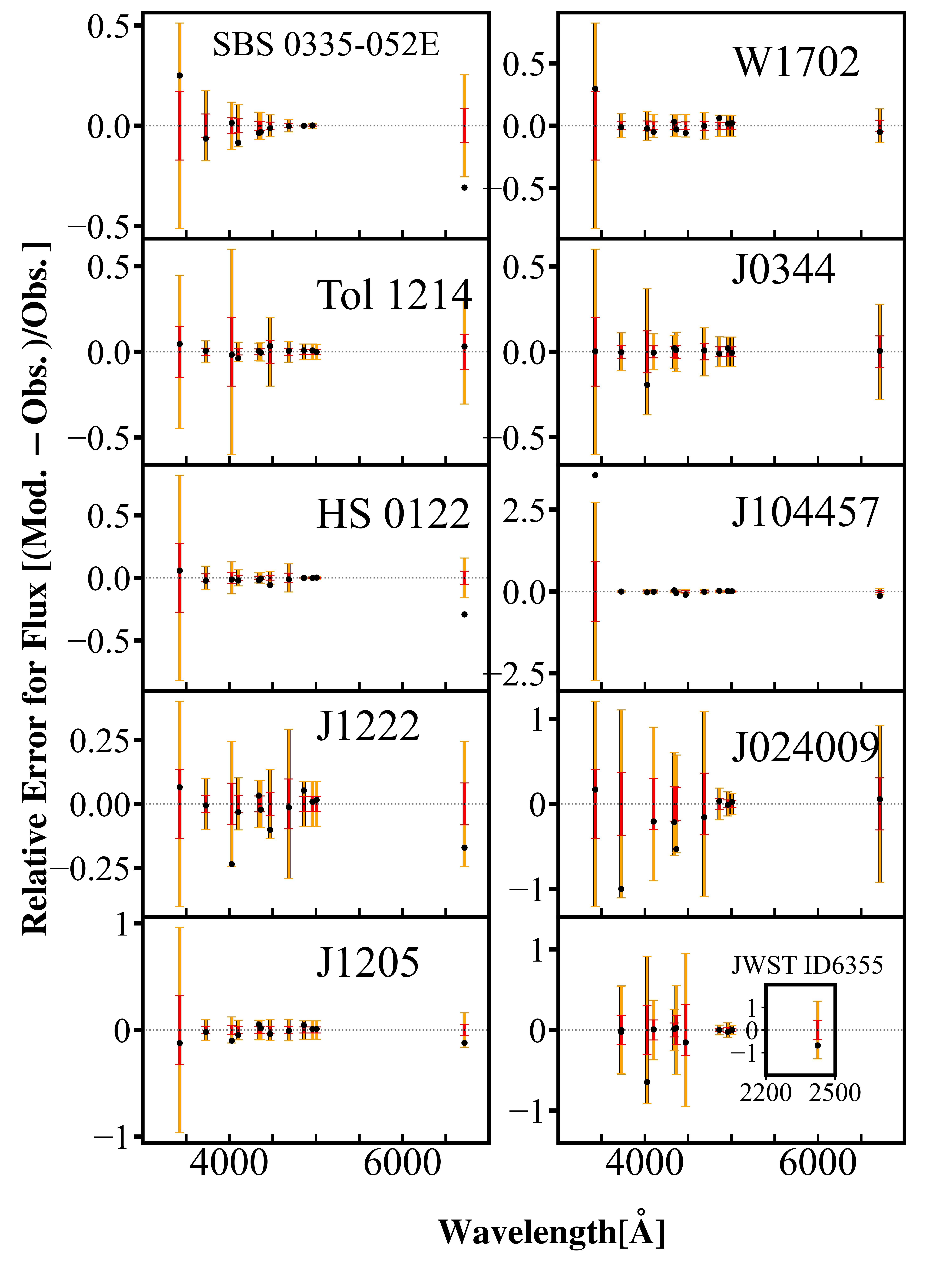}
    \caption{Differences of the best-estimate model line fluxes from the observed line fluxes for our ten galaxies. The vertical axes denote the differences defined in the main text. The best-estimate model fluxes are reproduced from the best-fit parameters. The red (orange) lines represent the 1$\sigma$ (3$\sigma$) errors of the observed fluxes.
    The inset panel in the bottom right panel shows the result of the [Ne {\sc iv}]/[Ne {\sc iii}] emission line ratio of ID 6355.
    }
    \label{match_diagram}
\end{figure*}

We test our best-fit parameters by comparing the observed flux measurements with model fluxes calculated from the best-fit parameters.
We define the relative residual fluxes by $(F_{\lambda, {\rm{mod}}} - F_{\lambda, {\rm{obs}}})/F_{\lambda, {\rm{obs}}}$,
and present the relative residual fluxes in Figure \ref{match_diagram}.
Figure \ref{match_diagram} indicates that the best-fit models reproduce almost all the observed emission lines within the 
3 sigma levels with an exception of the [Ne {\sc v}]/[Ne {\sc iii}] of J104457.
This exception is probably because J104457 have low S/N ($\sim 1$) for [Ne {\sc v}]/[Ne {\sc iii}] compared to other galaxies (S/N $\gtrsim $ 3). 
Hereafter, we conduct the same analysis for J104457 as other galaxies just for the presentation purpose.

\subsection{Best-Fit Ionizing Spectra}
\label{sec:best-fit_SED}
\begin{figure*}
    \centering
    \includegraphics[width=18cm]{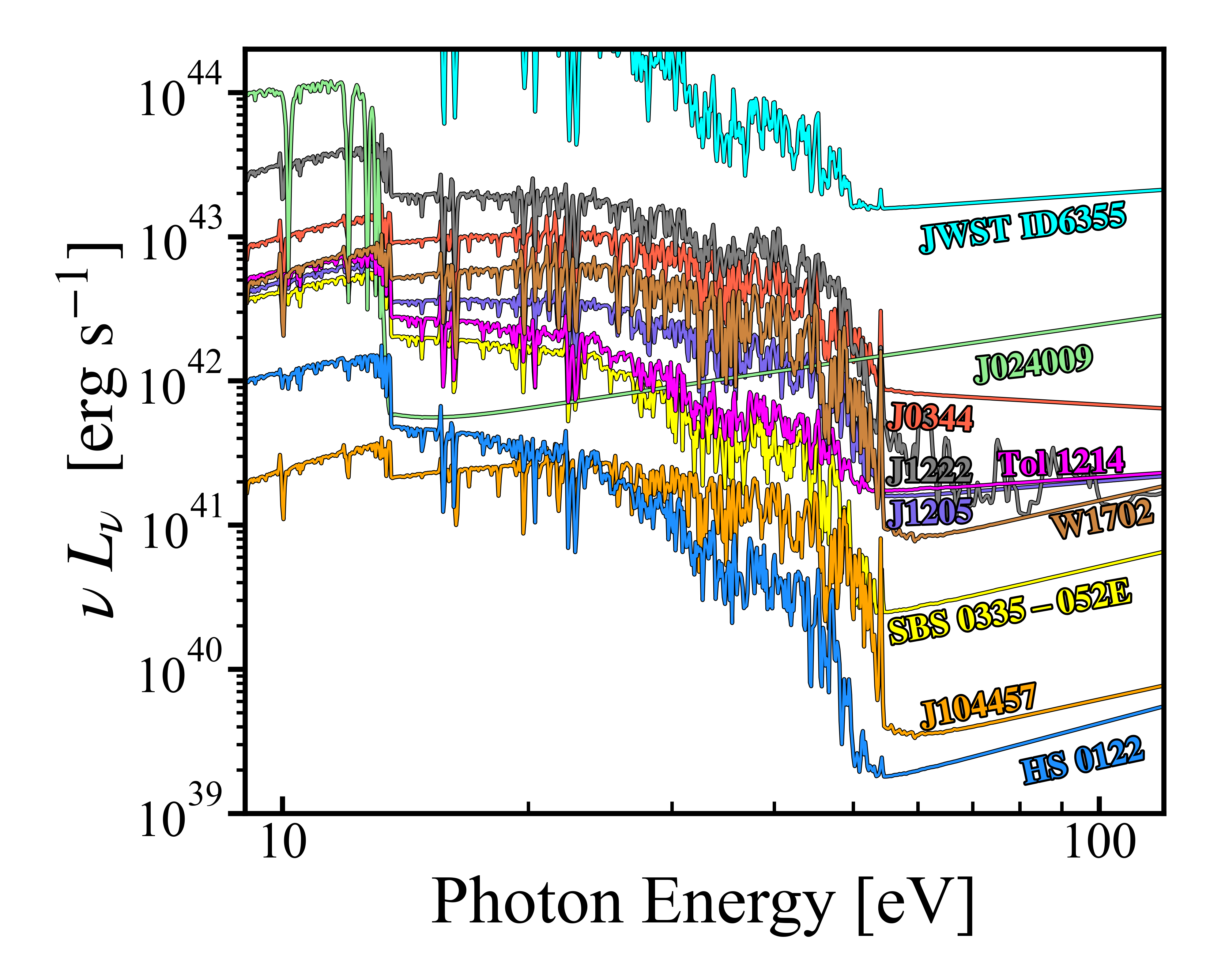}
    \caption{Ionizing spectra estimated at 13.6--100 eV for ten dwarf galaxies. All the ionizing spectra show prominent power-law continua in the 55-100 eV range.}
    \label{SED_EUV}
\end{figure*}

We determine the ionizing spectrum shapes $F_\nu$ with the best-fit parameters and Equations (\ref{eq:defin_spectra}) and (\ref{eq:power-law}).
We then calculate the luminosity $L_{\nu}$ from $F_\nu$ with a conversion factor $A$.
\begin{equation}
    L_{\nu} = A \ F_{\nu}.
\end{equation}
$A$ is obtained for each dwarf galaxy with a relation between the H$\beta$ luminosity $L({\rm{H}\beta})$ and the number of hydrogen ionizing photons produced per second given by \cite{2010ApJ...724.1524O}.
\begin{equation}
    A\times \int^{h\nu= \infty}_{h\nu = 13.6 {\rm eV}} \frac{F_{\nu}}{h\nu}\ {\rm{d}}\nu = \frac{L({\rm{H}}\beta)}{4.78\times 10^{-13}}.
\end{equation}
Here we assume escape fraction of ionizing photons $f_{\rm esc}^{\rm ion}$ as $f_{\rm esc}^{\rm ion}=0$. 
The $L({\rm{H}\beta})$ values are calculated from $F({\rm{H}\beta})$ presented in Table \ref{table:sample_properties}.
In Figure \ref{SED_EUV}, we present $L_{\nu}$ as a function of photon energy for all the galaxies. All the spectra of the galaxies show prominent power-law continua in the EUV range of 55--100 eV. 
The SED of local galaxies and ID 6355 are constrained in the range of $\sim 13.6 - 100$ and  $\sim 13.6 - 64$ eV, respectively.
 
We define the EUV luminosity $L_{\rm{EUV}}$ of the power-law continuum with the given parameters of $\alpha_{\rm X}$, $t$, and $a_{\rm{mix}}$. 
\begin{equation}
    L_{\rm{EUV}} = \int^{h\nu = {\rm 100 \ eV}}_{h \nu = {\rm 55 \ eV}} A \ C_{\rm{mix}}(a_{\rm{mix}}, t, \alpha_{\rm X}) \ P(\nu, \alpha_X) \ {\rm d}\nu .
\end{equation}
Here, $C_{\rm{mix}}$ is calculated with Equation (\ref{eq:C_mix}).
We define the EUV power-law index $\alpha_{\rm EUV}$ with $\alpha_{\rm EUV} =\alpha_{\rm{X}}$.

\begin{figure*}
    \centering
    \includegraphics[width=18cm] {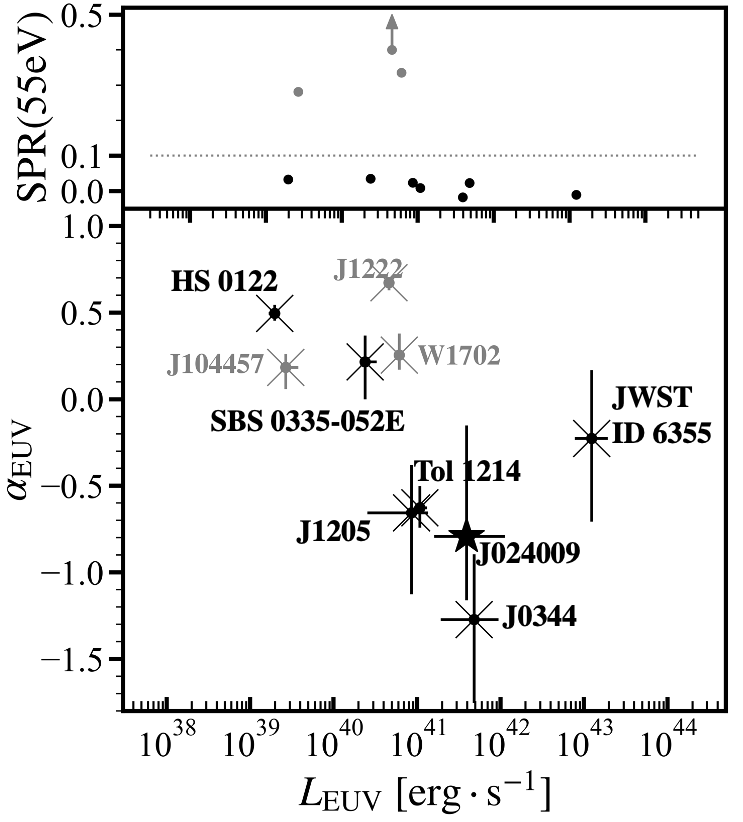}
    \caption{Bottom: EUV spectral slope as a function of EUV luminosity for ten dwarf galaxies. Top: Ratios of stellar flux to the power-law flux at 55 eV. 
    The error bars are defined by the values of 16 and 84 percentile of the extracted parameter sets.
    Criterion of stellar radiation to power-law continua ratio (0.1) are plotted as horizontal dots. 
    The dwarf galaxies with larger value of stellar radiation to power-law continua ratio is plotted in grey in the both panels.}
    \label{fig:LEUV_aEUV_without_model}
\end{figure*}

We extract the last 10 steps of the sampled parameter sets and calculate $\alpha_{\rm EUV}$ for all the extracted parameters. 
In some galaxies, the extracted $\alpha_{\rm EUV}$ values exhibited multiple peaks, including a primary peak associated with the best-fit parameters.
We additionally extract parameter sets in the primary peak for SBS 0335-052E, HS 0122, J1222, J1205, W1702, J104457, Tol 1214, J0344, and ID 6355 and calculate $L_{\rm{EUV}}$ and $\alpha_{\rm{EUV}}$ with the extracted parameter sets.
Figure \ref{fig:LEUV_aEUV_without_model} presents $\alpha_{\rm{EUV}}$ as a function of $L_{\rm{EUV}}$ for all our dwarf galaxies. 
For ID 6355, we extrapolate the power-law component of the best-fit spectrum up to 100 eV and calculate the $L_{\rm{EUV}}$.
We define the stellar to power-law ratio at 55 eV (SPR(55 eV)) as
\begin{equation}
    \rm SPR(55\ eV) = \frac{F_{\rm S, 55 \ eV}}{F_{\rm P, 55\ eV}}, 
\end{equation}
where $F_{\rm S, 55\ eV}$ and ${F_{\rm P, 55 \ eV}}$ are flux densities of stellar and power-law components at 55 eV of the best-fit spectrum, respectively. 
We calculate SPR(55 eV) for all the galaxies and show the SPR(55 eV) values at the top panel in Figure \ref{fig:LEUV_aEUV_without_model}.
The SPR(55 eV) distribution suggests that there are two different groups of galaxies.
Because we constrain the power-law component with two points, 54 eV (He {\sc ii}) and 97 eV ([Ne {\sc v}]), 
spectra with prominent stellar component would give additional uncertainties. 
We remove the group of galaxies, {
J104457, J1222, and W1702,} having prominent stellar contamination at 55 eV with criterion of ${\rm SPR(55\ eV)} > 0.1$. The removed galaxies are plotted in grey in Figure \ref{fig:LEUV_aEUV_without_model}.

\begin{figure*}
    \centering
      \includegraphics[width=18cm]{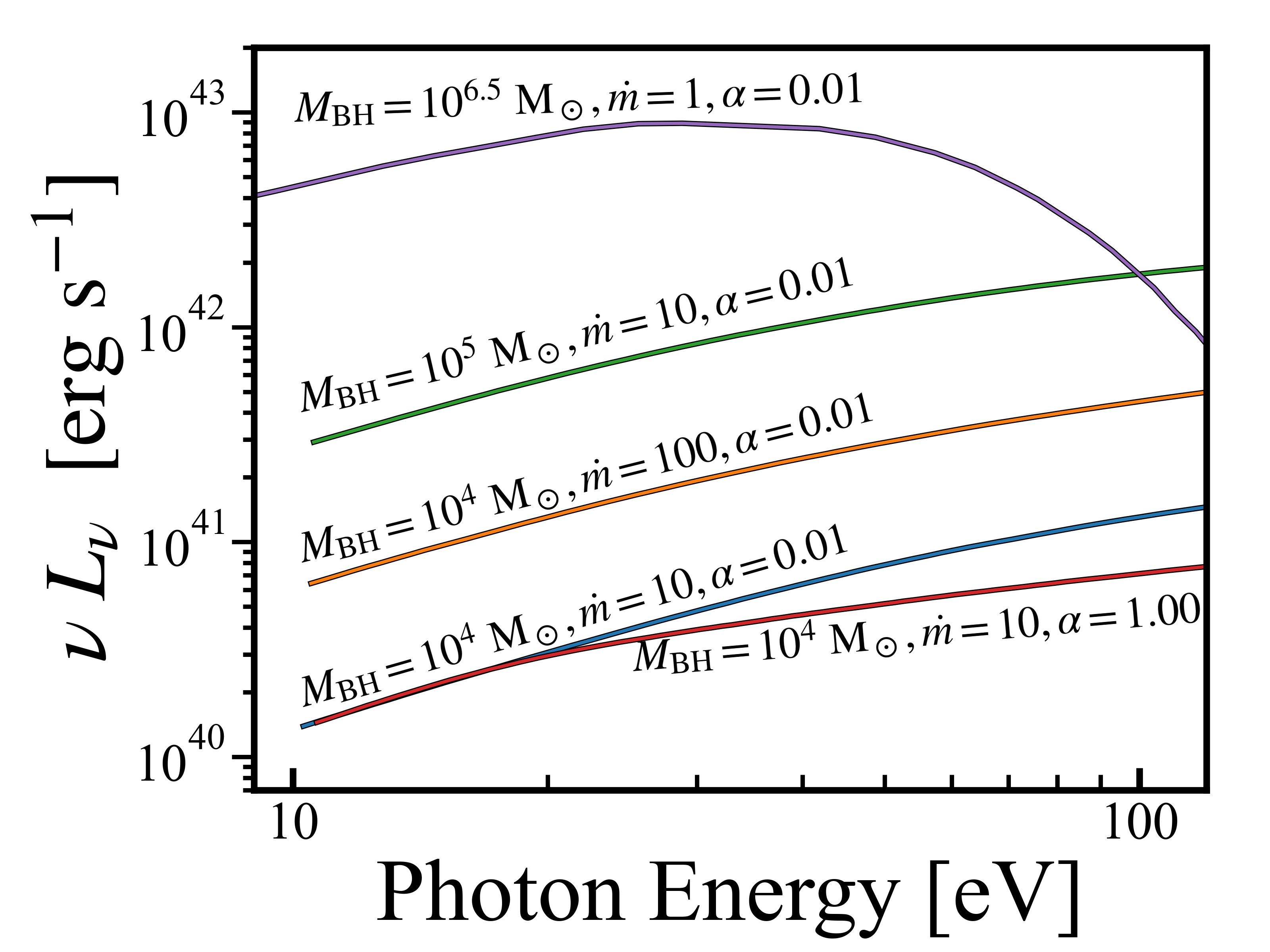}
    \caption{Examples of the K03 model spectra. Three parameters $M_{\rm BH}$, $\dot{m}$, and $\alpha$ 
    determine the BH accretion disk size (for a given temperature), temperature distribution, and the influence of electron scattering of the
    K03 models, 
    changing the spectral shape.
    When $M_{\rm BH}$ increases (with a fixed $\dot{m}$), the size of an accretion disk increases and the temperature decreases, changing the spectra brighter and redder. 
    When $\dot{m}$ increases, the temperature of an accretion disk increases, leading to brighter and bluer spectra. 
    When $\alpha$ increases, the effect of inverse Compton scattering increases, making spectra flatter (i.e., smaller $\alpha_{\rm EUV}$).
    Details are explained in K03. }
    \label{fig:SED_K03}
\end{figure*}

\section{Discussion}
\label{sec:discussion}
\subsection{Comparing Our Results with BH Accretion Disk Models}

We compare $\alpha_{\rm{EUV}}$ and $L_{\rm{EUV}}$ values (Section \ref{sec:best-fit_SED}) with BH accretion disk models given by \citeauthor{2003ApJ...593...69K} (\citeyear{2003ApJ...593...69K}; hereafter K03).
We estimate the set of BH masses and accretion rates that explain both $\alpha_{\rm{EUV}}$ and $L_{\rm{EUV}}$ values, assuming that the power-law continua in Section \ref{sec:ionizing_spectra} originate from BH accretion disks within physical parameters available in K03 models.

\subsubsection{K03 Models}
\label{sec:K03_LEUV_aEUV}
  The K03 models are accretion disk models 
  (within $2 \ \times 10^4$ Schwarzschild radii from the central BH)
  including effects of electron scattering and the relativistic correction, predicting spectra in the wavelength from the far UV to X-ray bands.
  These models have parameters of BH mass $M_{\rm{BH}}$, viscosity $\alpha$, and accretion rate $\dot{m}$ ranging in
  $ M_{\rm{BH}} = 10^2-10^{6.5}\ M_\odot$, $\alpha = 0.01-1$, and $\dot{m} = 1-1000$, respectively. 
  Model spectra with $M_{\rm BH} = $
1--$10^5\ M_{\odot}$ have been used to investigate
bright X-ray sources (e.g., \citealt{2010ApJ...722..760Y},
\citealt{2012ApJ...752...34G}).
The accretion rate parameter range covers {a physically reasonable range} from sub-Eddington ($\dot{m} < 16$) to super-Eddington accretion. 
Examples of the K03 model spectra are plotted in Figure \ref{fig:SED_K03}.

\begin{figure*}
    \centering
    \includegraphics[width=18cm]{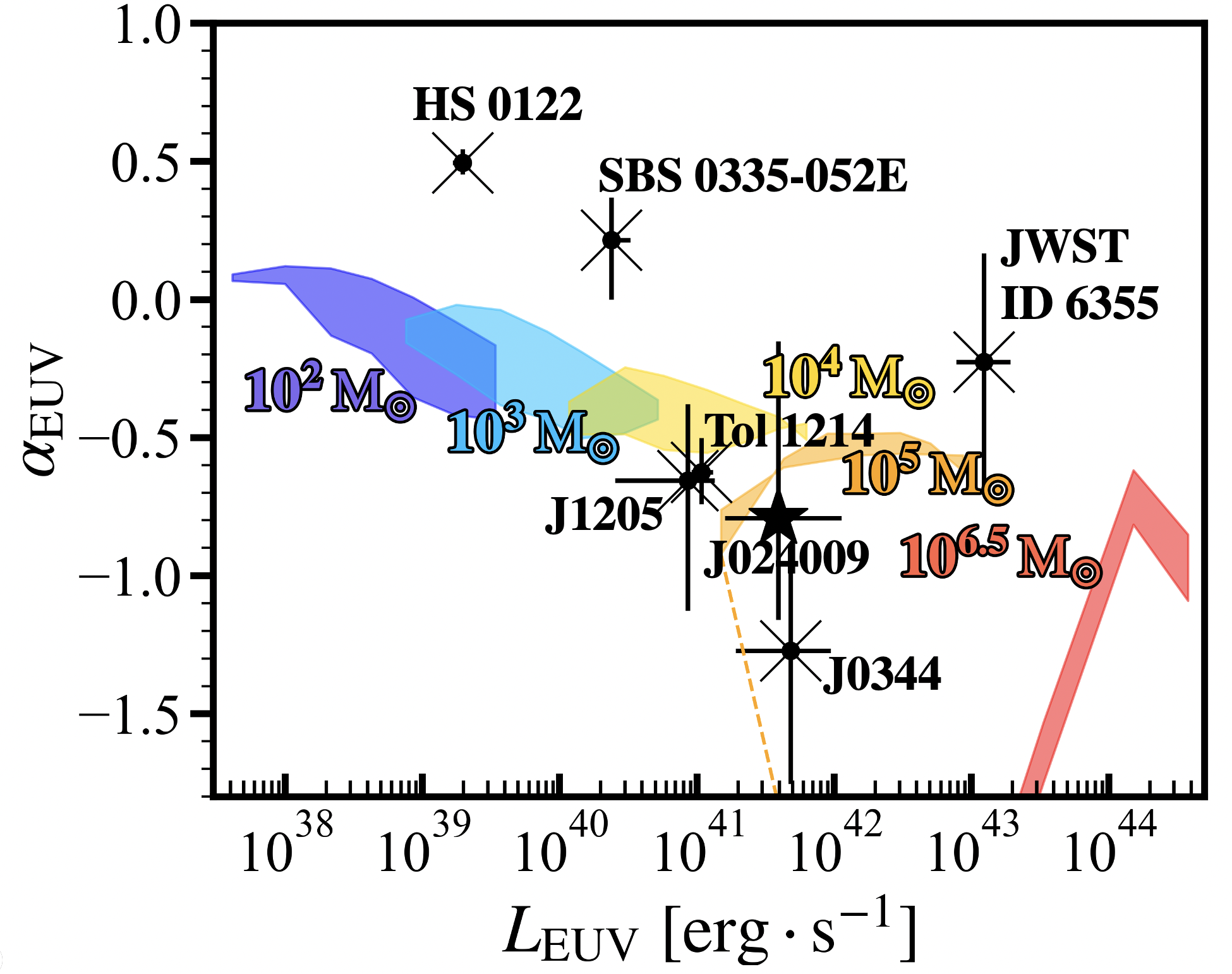}
    \caption{Spectral slope of power-law continua and EUV luminosities of BH accretion disk models and reproduced ionizing spectra for seven dwarf galaxies (symbols). Color shaded regions represent where BH accretion 
    represent the BH disk model results for the accretion rate and viscosity parameter varying in the range of 
    $\dot{m}=$1-1000 and $\alpha = 0.01-1$.  
    Difference in color shows difference in BH masses of the K03 models. Each line shows how a position on the graph change when accretion rate changes. 
    Upper and lower line represent model with different viscosity. 
    We filled between the two lines. 
    For $M_{\rm BH} = 10^{6.5}\ M_\odot$, we used the spectra for $\alpha$ of 0.1 (with $\dot{m}$ from 1 to 1000; Fig.12 of K03), and then broaden the spectral slope by $\pm 0.1$ to account for the wider viscosity range (from 0.01 to 1; Fig.9 of K03).
    The orange dashed line connect the $\alpha_{\rm EUV}$ and $L_{\rm EUV}$ values of $(M_{\rm BH}, \dot{m}, \alpha) = (10^5 M_\odot, 1, 1)$ and $(10^{6.5} M_\odot, 1, 0.1)$.}
    \label{fig:K03_and_galaxies}
\end{figure*}

  We determine the K03 models' EUV luminosities $L_{\rm{EUV, K03}}$ and power-law indexes $\alpha_{\rm{EUV,K03}}$. 
  We calculate $L_{\rm{EUV, K03}}$, integrating the K03 model spectra over 55--100 eV. 
  We estimate $\alpha_{\rm{EUV,K03}}$ by fitting a power law to the K03 model spectra in the 55--100 eV range using $\tt{numpy.polyfit}$. 
  Figure \ref{fig:K03_and_galaxies} presents $\alpha_{\rm{EUV,K03}}$ and $L_{\rm{EUV, K03}}$ values of the K03 models in the given parameter ranges.

\subsubsection{Black Hole Mass Estimates}
\label{sec:bh_mass_estimate}
  In Figure \ref{fig:K03_and_galaxies}, we compare the K03 models ($\alpha_{\rm{EUV,K03}}$ and $L_{\rm{EUV, K03}}$; Section \ref{sec:K03_LEUV_aEUV}) with observations ($\alpha_{\rm{EUV}}$ and $L_{\rm{EUV}}$; Section \ref{sec:best-fit_SED}).
  First, we test our BH mass estimation method with J024009 that has an IMBH with $M_{\rm BH}=10^{5.75} \ M_\odot$ and super-Eddington accretion ($\dot{m} > 16$) measured by optical spectroscopic observations \citep{2011ApJ...739...28X}. We show J024009 with the star mark in Figure \ref{fig:K03_and_galaxies}. J024009 falls on the models of $M_{\rm BH}=10^5 \ M_\odot$ suggesting that the BH mass of J024009 is $M_{\rm BH}\sim 10^5 \ M_\odot$. 
  Also an accretion rate is consistent with super-Eddington accretion within errorbars. 
  We thus regard that our BH mass estimation method is applicable to the galaxies 
  with BH accretion disks within physical parameters available in K03 models. 

  Figure \ref{fig:K03_and_galaxies} indicates that 
  the $L_{\rm EUV}$ and $\alpha_{\rm EUV}$ of SBS 0335-052E (J1205 and Tol 1214) agrees with the models with $M_{\rm BH}=10^{3-4}\ M_\odot$ ($M_{\rm BH}=10^{4-5}\  M_\odot$) within the $2\sigma$ level and therefore the BH mass is estimated to be $M_{\rm BH}\sim 10^{3-4}\ M_\odot$ ($M_{\rm BH}\sim 10^{4-5}\ M_\odot$). 
  J0344 is placed outside the regions covered by the models. 
  However, J0344 is placed on or slightly above the orange dashed line connecting the data point of $(M_{\rm BH}, \dot{m}, \alpha) = (10^5 M_\odot, 1, 1)$ and $(10^{6.5} M_\odot, 1, 0.1)$. This suggest that J0344 is explained by BH accretion disk model with BH mass between $10^5$ and $10^{6.5} M_\odot$ with sub-Eddington mass accretion. This suggests that J0344 harbors a massive BH with a mass of $M_{\rm BH} \sim 10^5-10^{6.5}\ M_\odot$.
  The $L_{\rm EUV}$ and $\alpha_{\rm EUV}$ of ID 6355 are consistent with those of $M_{\rm BH}\sim10^5 \ M_{\rm \odot}$ BH accretion disk models within errorbars. This suggest that the ID 6355 is harboring a BH with a mass of $M_{\rm BH}\sim10^5 \ M_{\rm \odot}$.


 \begin{figure*}
    \centering
    \includegraphics[width=18cm]{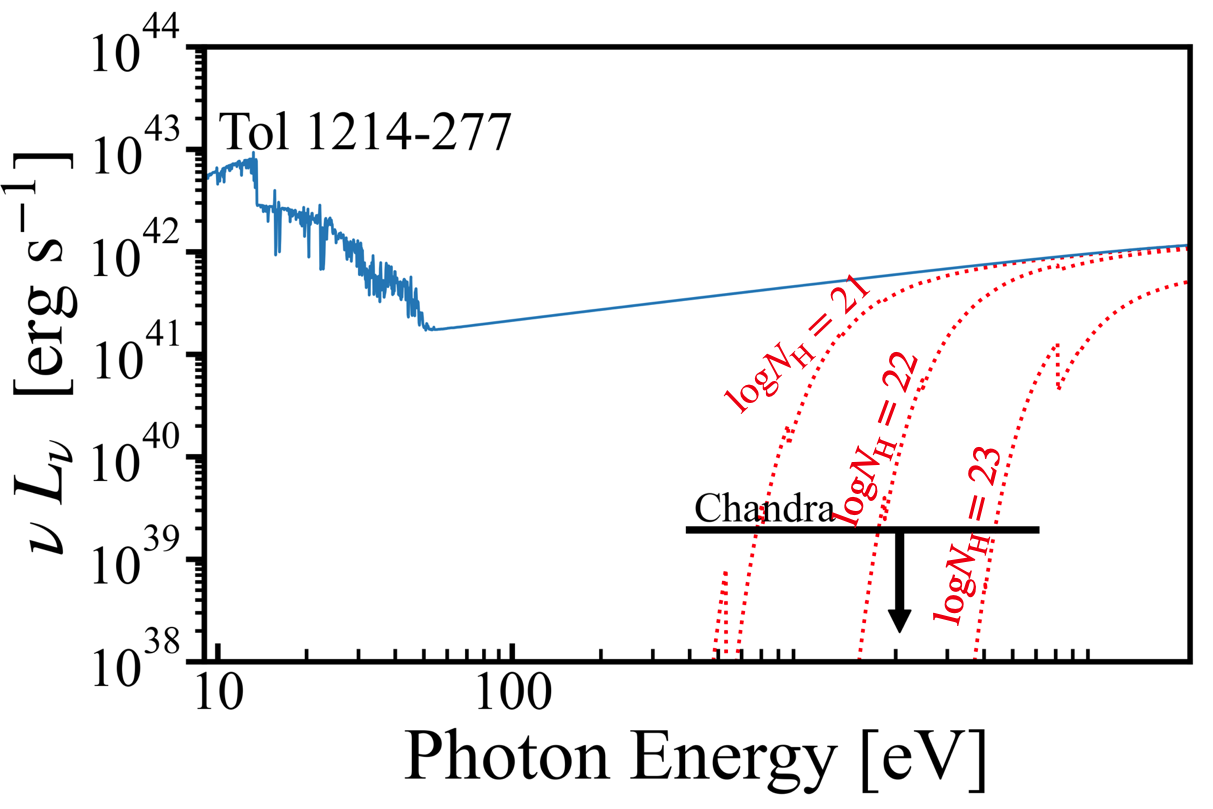}
    \caption{{Comparison between the X-ray observation and extrapolation of the best-fit SED of Tol 1214. 
    The black line and arrow denote the upper limit of X-ray luminosity given by Chandra observations.
    The red dotted lines indicate net transmitted spectra assuming hydrogen column densities of $\log (N_{\rm H}/{\rm cm^{-2}})=$21,22, and 23.
    }
    }
    \label{fig:X-ray}
\end{figure*}

\begin{figure*}
    \centering
    \includegraphics[width=18cm]{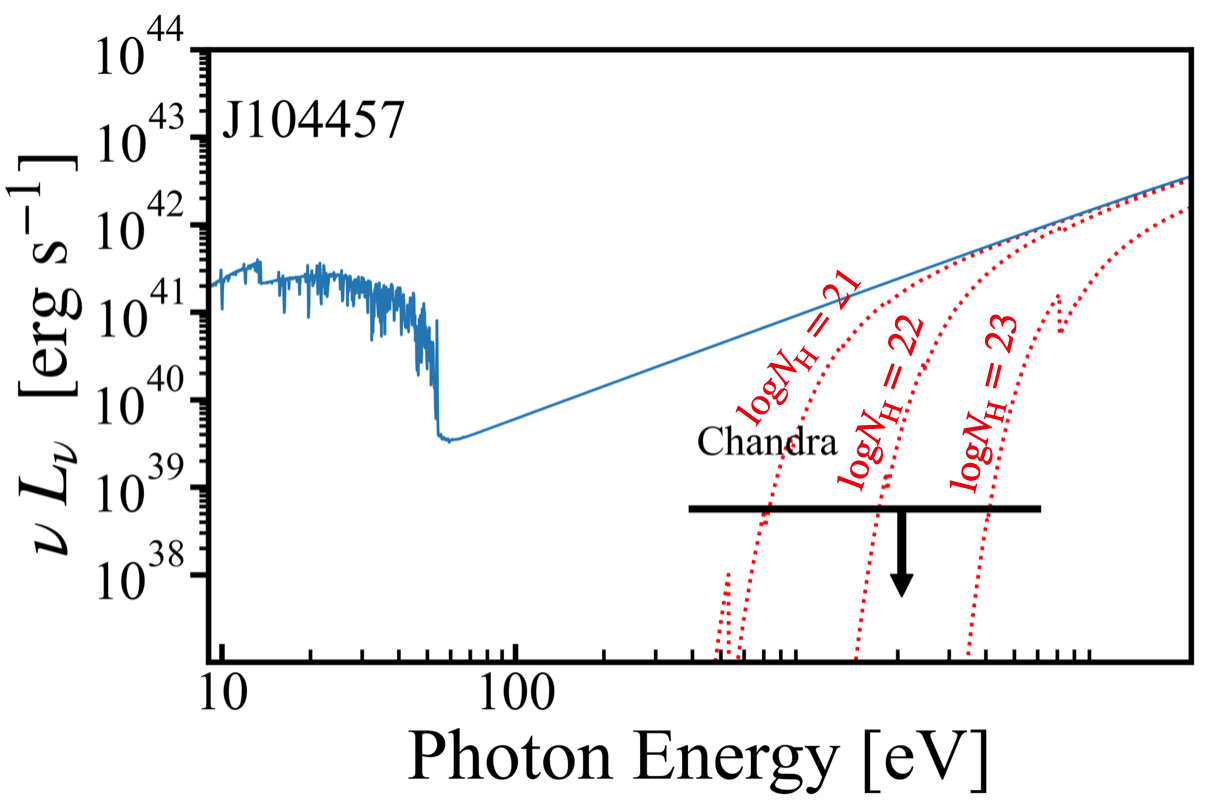}
    \caption{{Same as in Fig \ref{fig:X-ray}, but for J104457. }}
    \label{fig:X-ray_J104457}
\end{figure*}

 \begin{figure*}
    \centering
    \includegraphics[width=18cm]{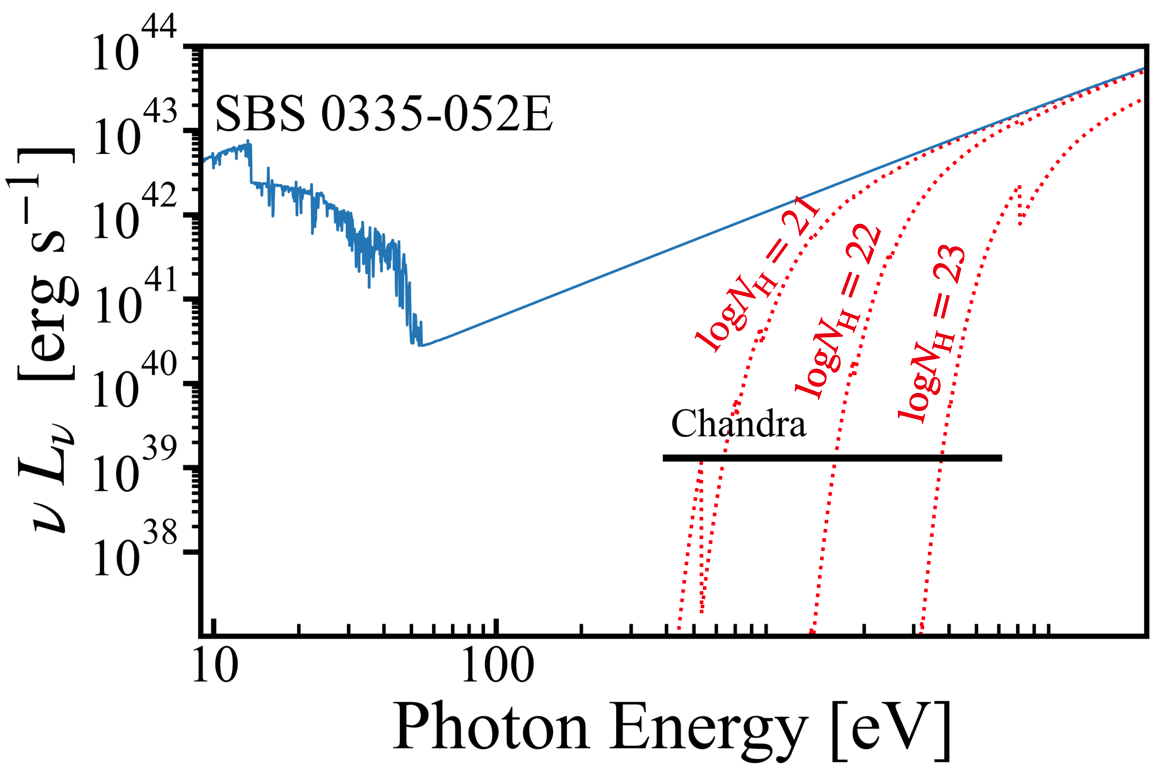}
    \caption{{Comparison between the X-ray observation and extrapolation of the best-fit SED of SBS 0335-052E. 
    Black line denote the X-ray luminosity given by Chandra observations \citep{2004ApJ...606..213T}.
    The red dotted lines indicate net transmitted spectra assuming hydrogen column densities of $\log (N_{\rm H}/{\rm cm^{-2}})=$21,22, and 23.
    }}
    \label{fig:X-ray_ID50_v2}
\end{figure*}

\subsection{BH Mass to Stellar Mass Relation}

\begin{figure*}
    \centering
    \includegraphics[width=18cm]{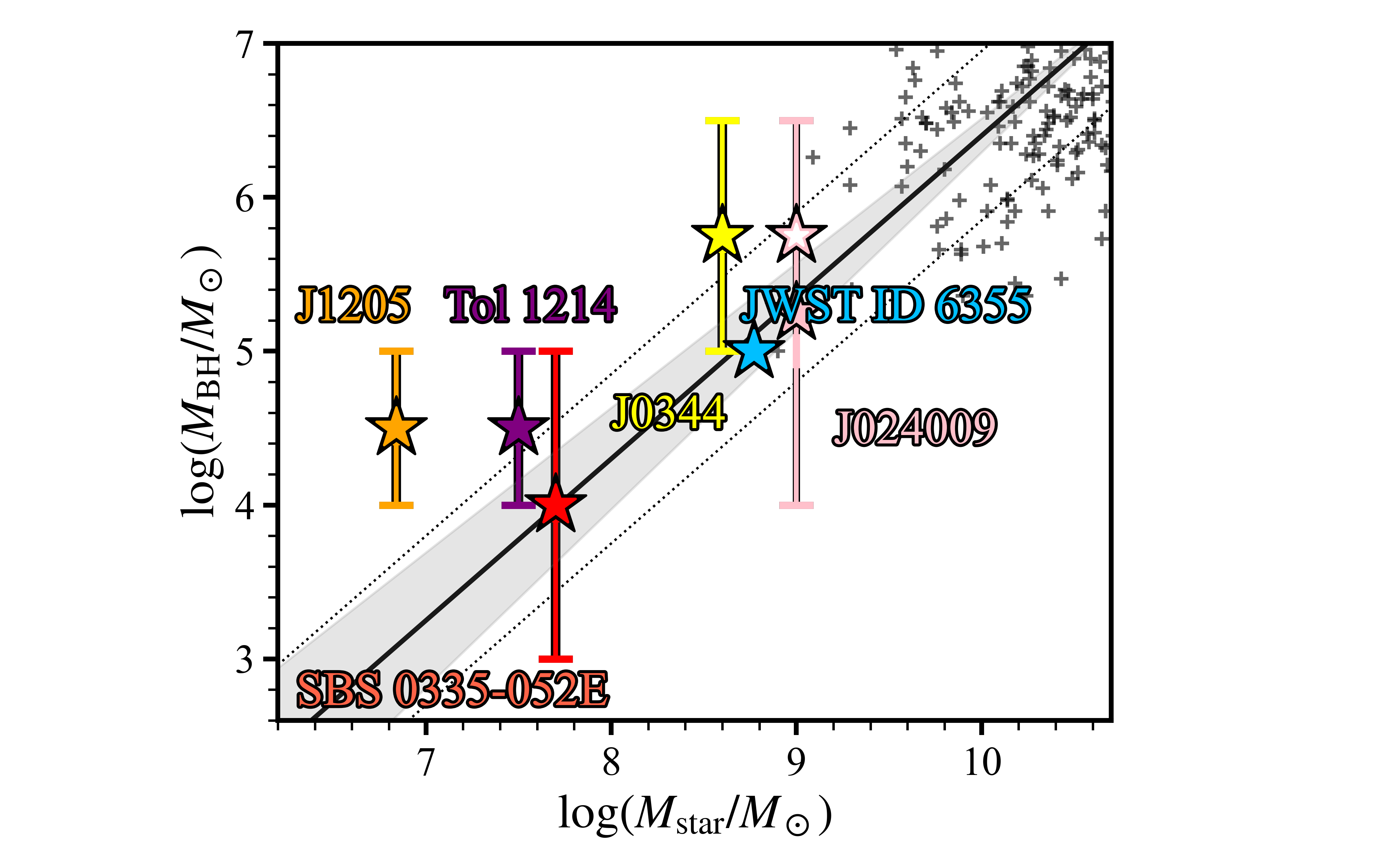}
    \caption{Relation between black hole mass $M_{\rm BH}$ and host galaxy stellar mass $M_{\rm star}$ for the seven dwarf galaxies. Color bars and star marks represent the BH mass ranges of the K03 models that are consistent with the $L_{\rm EUV}$ and $\alpha_{\rm EUV}$ values of the six dwarf galaxies within 2$\sigma$ levels.
    The star marks are plotted at the centers of the BH ranges. The pink open star represents the BH mass of J024009 given by \cite{2011ApJ...739...28X} derived from the H$\alpha$ broad line. The grey cross symbols represent the black hole and stellar masses of local dwarf galaxies provided by \cite{2015ApJ...813...82R}. The black solid line and the shaded region indicate the local relation in the range of $M_{\rm BH} = 10^5 - 10^{8.5}\ M_\odot$ given in \cite{2015ApJ...813...82R}. The black dotted lines show the sum of intrinsic scatter and measurement uncertainties of the local relation.}
    \label{fig:BH_star}
\end{figure*}

We plot the stellar and BH masses of the 6 galaxies with those of local dwarf galaxies in Figure \ref{fig:BH_star}.
The stellar masses of J0344, Tol 1214-277, and J024009 are calculated with the relation of absolute $i$-band magnitudes and stellar masses given in \cite{2021ApJ...918...54I}. 
The galaxies fall on the line or above the extrapolation of the local relation given in \cite{2015ApJ...813...82R}, suggesting that the BH mass to stellar mass ratio is the same or larger in the low mass range than in the high mass range.

\subsection{{Emission Line Profiles}}

\begin{table}
\begin{center}
  \caption{{BH masses estimated from H$\alpha$ broad lines.}}
  \label{table:M_BH}
  \centering
  \begin{tabular}{lcc}
    \hline
   Name    & $M_{\rm BH}$  &  $M_{\rm BH}$ \\
    &   H$\alpha$ broad line &  Our methods  \\
    &   $M_\odot$ & $M_\odot$ \\
    \hline
    Tol 1214  & $\gtrsim 1.9\times10^{4}$  & $\sim 10^{4-5}$ \\
    SBS 0335-052E & $\lesssim 1.4 \times 10^{8}$ & $\sim 10^{3-5}$\\
    J1205  & $1.5\times 10^{5}$ & $\sim 10^{4-5}$ \\
    J0344  & $3.8\times10^{5}$ & $\sim 10^5$\\
    \hline
\end{tabular}
  \end{center}
\end{table}

\begin{table}
\begin{center}
  \caption{{Emission line profiles of SBS 0335-052E and HS 0122.}}
  \label{table:FWHM}
  \centering
  \begin{tabular}{lcccc}
    \hline
   Name  &  & & FWHM  & \\
    & & [Ne {\sc v}]$\lambda$3426  & He {\sc ii}$\lambda$4686 &H$\beta$  \\
    \hline
    SBS 0335-052E  &  [${\rm \AA}$] & 1.4 & 1.3 & 1.3 \\
    &     [$\rm km \ s^{-1}$] &  283 & 192 & 186 \\
    HS 0122 &  [${\rm \AA}$] & 1.8 & 1.3 & 1.2 \\
      &   [$\rm km \ s^{-1}$] & 370  & 200 & 170 \\
    \hline
\end{tabular}
  \end{center}
\end{table}

\begin{table*}
\begin{center}
  \caption{{[Ne V] luminosities of dwarf galaies.}}
  \label{table:L_NeV}
  \centering
  \begin{tabular}{lcc}
    \hline
   Name & [Ne V]$\lambda$3426 luminosity &  Reference \\
   & log $L$([Ne V]$\lambda$3426)/ erg s$^{-1}$ (dex) & \\
    \hline 
    SBS 0335-052E & $38.1 \pm 0.1$  & (This work) \\
    HS 0122+0743 & $ 37.5 \pm 0.1$ & (This work) \\
    J104457 & 36.6 $\pm$ 0.7 &  \cite{2021ApJ...922..170B}\\
    J1222+3602 & 39.6 $\pm$ 0.1 & \cite{2021MNRAS.508.2556I} \\
    W1702+18 & 38.2 $\pm$ 0.2 & \cite{2021MNRAS.508.2556I} \\
    Tol 1214-277& $38.9 \pm 0.1$ & \cite{Izotov2004} \\
    J1205+4551 & $38.5 \pm 0.1$ & \cite{2021MNRAS.508.2556I}\\
    J0344-0106& $ 39.3 \pm 0.1$ & \cite{2021MNRAS.508.2556I} \\
    J024009.10+010334.5 & $40.6 \pm 0.3$ & (This work) \\
    \hline
    SN2010jl & 38.5 & (This work)\\
    SDSS DR8 & 41.21$\pm$0.29 & \cite{2022ApJ...936..140R}\\
    AGNs (z$\sim$1) & $>$41.5 & \cite{2023ApJ...948..112C}\\
    AGNs (z$\sim$0) & 37$-$42 & \cite{2010Gilli}\\
    \hline
    \label{FWHM}
\end{tabular}
  \end{center}
\end{table*}

{
AGNs often show broad H$\alpha$ lines originating from the surrounding broad line regions, and the BH masses can be estimated from the strength and the width of the H$\alpha$ lines \citep{2005ApJ...630..122G}. 
}
{
There are H$\alpha$ and H$\beta$ broad lines with FWHM $\gtrsim 1000 \ {\rm km \ s^{-1}}$ reported in some of the dwarf galaxies in previous studies (e.g. \citealt{2021MNRAS.508.2556I,2017MNRAS.471..548I,2005ApJS..161..240T}). 
We investigate the H$\alpha$ and H$\beta$ broad lines of galaxies shown in Figure 7: HS 0122, Tol 1214, SBS 0335-052E, J1205, and J0344. 
We find a report of the broad line of Tol 1214 in \cite{2021MNRAS.508.2556I, 2004A&A...421..539I}. The lower limit of the H$\beta$ FWHM of Tol 1214 is $\gtrsim 900 \ {\rm km \ s^{-1}}$ and broad line to narrow line flux ratios are  $\sim$ 1-2 $\%$. For SBS 0335-052E, the H$\alpha$ broad line is identified in \cite{2023arXiv230403726H} who place the BH mass upper limit of  $M_{\rm BH} < 1.4\times 10^{8}\  M_\odot$.
For HS 0122, we search for H$\alpha$ or H$\beta$ broad lines in the LRIS spectra reduced in Section \ref{sec:data_reduction}. 
We find possible broad H$\alpha$ and H$\alpha$ lines in HS 0122, while these broad lines may originate from the instrumental broadening.
For J1205 and J0344, there are reports of H$\alpha$ broad lines in \cite{2021MNRAS.508.2556I,2017MNRAS.471..548I}. 
For J0344, only the velocity dispersion of H$\alpha$ broad line is given and the line flux is not shown in \cite{2021MNRAS.508.2556I}. 
For J0344, we assume the broad line flux to the whole H$\alpha$ flux {ratio} is 0.01.}

{We estimate BH masses of Tol 1214, SBS 0335-052E, J1205, and J0344 from the velocity dispersions and fluxes of H$\alpha$ broad line, using the equation (6) of \cite{2005ApJ...630..122G}. We compare the estimated BH masses with those obtained from our models in Table \ref{table:M_BH}.  
{The BH masses of J1205, Tol 1214, and J0344 calculated from the H$\alpha$ broad lines are similar to the BH masses estimated from our methods.
For SBS 0335-052E, the black hole mass derived from H$\alpha$ broad lines is larger than that derived from our methods. 
We disscuss the difference in \cite{2023arXiv230403726H}, including discussion of SEDs and outflows. 
}} 

{
  For all of our sample galaxies with the LRIS spectra, SBS 0335-052E and HS 0122, we have measured the FWHMs of H$\beta$, He {\sc{ii}}$\lambda 4686$, and [Ne {\sc{v}}]$\lambda 3426$ emission whose ionization potentials range widely, 13.6, 54.4, and 97 eV, respectively, evaluating the FWHMs with a Gaussian function. 
  We show the FWHM measurements in Table \ref{table:FWHM}.}
{  
We find that the FWHMs are comparable to instrumental broadening, and are smaller than those of weak broad lines with FWHM $\gtrsim$ 1000 km/s found in strong emission lines (e.g.  [O {\sc iii}]$\lambda \lambda$4959,5007, H$\alpha$).}

\subsection{{Comparison with X-ray observations}} 

{We compare the X-ray observation with the best-fit spectra of dwarf galaxies and some accretion disk models.
We find X-ray observational resutls for Tol 1214, SBS 0335-052E, and J104457 in the litearture and Chandra Data Archive.
X-ray emission are not detected in Tol 1214 and J104457, but SBS 0335-052E. For Tol 1214 and J104457, we calculate the upper limits of the X-ray luminosities assuming the sensitivity limit of $4\times10^{-15}\ $erg s$^{-1}$\ cm$^{-2}$, and the flat spectra\footnote{\url{https://cxc.harvard.edu/cdo/about_chandra/}}}. 
The X-ray luminosity for SBS 0335-052E is {taken from} \cite{2004ApJ...606..213T}.

{We plot the X-ray luminosity measurements and upper limits for the three galaxies in Figures \ref{fig:X-ray}, \ref{fig:X-ray_J104457}, and \ref{fig:X-ray_ID50_v2}. 
We overplot the best-fit $\sim$13.6-100 eV spectra derived in Section \ref{sec:best-fit_SED} for each galaxy. To compare the best-fit ionizing spectra with the X-ray luminosity measurements and upper limits, we extrapolate the best-fit ionizing spectra, applying photoelectric absorption and Thomson scattering effects to the best-fit ionizing spectra with the photoelectric-absorption cross sections shown in Table 2 of \cite{1983ApJ...270..119M}.
We assume that hydrogen column densities of log $(N_{\rm H}/{\rm cm^{2}})$ = 21, 22, and 23 that cover both typical hydrogen scolumn densities reported in HMXBs log $(N_{\rm H}/{\rm cm^{2}})$ $\sim$ 21 and heavily obscured AGNs log $(N_{\rm H}/{\rm cm^{2}}$) $\sim$ 23.
We find that the best-fit ionizing spectra with a low hydrogen column density of log $(N_{\rm H}/{\rm cm^{2}})$= 21 are inconsistent with the X-ray luminosity measurements and upper limits, and that a high column density of log $(N_{\rm H}/{\rm cm^{2}})$ = 22 or 23 is required.
}



\subsection{{ {[Ne {\sc v}]$\lambda$3426 luminosities}}   }
{
Besides our sample of dwarf galaxies, [Ne {\sc v}]$\lambda$3426 lines are reported in some SNe and AGNs. 
We compare the [Ne {\sc v}]$\lambda$3426 luminosities with those of a supernova and AGNs.
}

{
We calculate the [Ne {\sc v}]$\lambda$3426 luminosities of the 9 dwarf galaxies with the [Ne {\sc v}]$\lambda$3426 line detection in our sample, and obtain the [Ne {\sc v}]$\lambda$3426 luminosities ranging in $\sim 10^{36-40}$ erg s$^{-1}$. These luminosities are presented in Table \ref{table:L_NeV}. 
We find [Ne {\sc v}]$\lambda$3426 luminosities of AGNs in three studies of \cite{2022ApJ...936..140R, 2023ApJ...948..112C,2010Gilli}. 
We find that the AGN luminosities of log $(L/{\rm erg \ s^{-1}})\gtrsim 41$ \citep{2022ApJ...936..140R, 2023ApJ...948..112C} and log $(L/{\rm erg \ s^{-1}}) \sim 37-42$ \citep{2010Gilli}. 
These [Ne {\sc v}]$\lambda$3426 luminosities of the AGNs are comparable to or larger than our dwarf galaxies.
}

{We compare [Ne {\sc v}]$\lambda$3426 luminosities of our sample galaxies with that of a supernova, SN 2010 jl \citep{2014ApJ...797..118F}.
We obtain a reduced optical spectrum of SN 2010 jl from wiserep database \citep{2012PASP..124..668Y}, 
and measure the [Ne {\sc v}]$\lambda$3426 luminosity to be $3.5\times10^{38}$ erg s$^{-1}$. This [Ne {\sc v}]$\lambda$3426 luminosity is comparable with our dwarf galaxies.
}

\subsection{Other high-ionization lines}
{In typical AGNs with high stellar masses and metallicities, [Fe {\sc v}]$\lambda$4227 and [Fe {\sc vii}]$\lambda$6087 lines are widely detected. 
We search the Fe high-ionization lines in the 10 dwarf galaxies. 
The [Fe {\sc v}]$\lambda$4227 lines are detected in J0344, J1205, W1702, Tol 1214, and SBS 0335-052E, and J104457 (\citealt{2021MNRAS.508.2556I,2021ApJ...922..170B}; This work).  
The [Fe {\sc vii}]$\lambda$6087 lines are detected in J1205 and SBS 0335-052E (\citealt{2021MNRAS.508.2556I}; This work).
We show the [Fe {\sc vii}]$\lambda$6087/[Fe {\sc v}]$\lambda$4227 flux {ratios} of J1205 and SBS 0335-052E in Table \ref{table:Fe}. 
To check whether the presence and absence of the emission lines are consistent with our model or not, 
we predict [Fe {\sc vii}]$\lambda$6087/[Fe {\sc v}]$\lambda$4227 from our best-fit models. 
We find that the best-fit model results are consistent with the observations within the $\lesssim$ 3 $\sigma$ uncertainties.}
{From the JWST spectra, we obtain [Ne {\sc iii}]$\lambda$3869/[Ne {\sc v}]$\lambda$3426 upper limit of $<9.3 \times 10^{-3}$. 
We predict the [Ne {\sc iii}]$\lambda$3869/[Ne {\sc v}]$\lambda$3426 ratio from the best-fit model as $8.7 \times 10^{-3}$. The non detection of the [Ne {\sc v}]$\lambda$3426 line in ID 6355 is consistent with the best-fit parameters.}

\begin{table*}
\begin{center}
  \caption{{Observed and predicted flux ratios of high-ionization lines of Fe in the dwarf galaxies.}}
  \label{table:Fe}
  \centering
  \begin{tabular}{lccccc}
    \hline
   Name  &  [Fe {\sc vii}]$\lambda$6087/[Fe {\sc v}]$\lambda$4227 & & Reference \\
     &  best-fit model prediction & observed  & \\
\hline
    SBS 0335-052E  &  0.112 & 0.0839 $\pm$ 0.024 &  (This Work)\\
    J1205          &  0.048 & 0.151  $\pm$ 0.029  & (\citealt{2021MNRAS.508.2556I})\\
    \hline
    \label{Fe}
\end{tabular}
  \end{center}
\end{table*}

\section{Summary}

We reconstruct the ionizing spectra of the dwarf galaxies in 13.6 - 100 eV range using $>$10 optical emission lines including faint high-ionization lines of He {\sc{ii}}$\lambda 4686$, [Ne {\sc{iv}}]$\lambda 2424$, and [Ne {\sc{v}}]$\lambda 3426$. 
We conduct deep optical spectroscopic observations for two dwarf galaxies classified as EMPGs with the Keck/LRIS spectrograph.
We make a total of 
the ten dwarf galaxies including EMPGs with detection of faint high-ionization lines 
from our observational data of the two dwarf galaxies, 
adding the eight dwarf galaxies from the literature.  
We derive the ionizing spectra at $13.6-100$ eV 
by the comparisons of the observed optical emission lines
and the photoionization models in the same manner as \citetalias{2022ApJ...930...37U}
with the two major improvements
to determine the high energy spectra in the EUV 
$\sim 55-100$ eV band. One improvement
is replacing blackbody spectra with the realistic 
stellar population spectra that affect the shapes of
the EUV spectra, while the other is
including the high ionization lines 
of [Ne {\sc{iv}}]$\lambda 2424$ and [Ne {\sc{v}}]$\lambda 3426$
whose ionization potentials are $\sim 60-100$ eV.
Our findings are listed below.

\begin{enumerate}
\item  For our ten galaxies, we derive the ionizing spectra over 13.6--100 eV that reproduce all of the observed emission line fluxes within $\lesssim 3\sigma$ errors. 
The ionizing spectra of the ten galaxies show prominent power-law radiation in the EUV band. 
We calculate power-law spectral properties of $L_{\rm EUV}$ and $\alpha_{\rm EUV}$, and find the anti-correlation for the ten galaxies.

\item As for the testing purpose, we compare $L_{\rm EUV}$ and $\alpha_{\rm EUV}$ of a known IMBH having a BH mass of $M_{\rm BH} = 10^{5.75}\ M_\odot$ and an approximate accretion rate of $\dot{m}\gtrsim 16$
(J024009; \citealt{2011ApJ...739...28X})
with those of the BH accretion disk models of K03.
We find that the IMBH agrees with the K03 models 
of $M_{\rm BH} = 10^5-10^{6.5}\ M_\odot$ and $\dot{m}=1-30$ on the $L_{\rm EUV}$-$\alpha_{\rm EUV}$ plane,
suggesting that the IMBH is explained by the
K03 model and that the BH mass of the IMBH 
is reproduced by the K03 model comparisons, {while we do not rule out other scenarios.}
We thus regard that this K03-model comparison method with the ionizing spectral properties of $\alpha_{\rm EUV}$ and $L_{\rm EUV}$ is applicable to galaxies with IMBHs.

\item We find that the three dwarf galaxies have ionizing spectra whose fluxes at 55 eV are contaminated significantly 
by the stellar radiation with SPR(55 eV)$>0.1$.
We further investigate the rest of the dwarf galaxies,
HS0122, SBS 0335-052E, Tol 1214, J1205, J0344, and ID 6355
by the K03-model comparison method.
In the $\alpha_{\rm EUV}$ and $L_{\rm EUV}$ plot of Figure \ref{fig:K03_and_galaxies}, five dwarf galaxies fall on the K03 models
of 
$M_{\rm BH} \sim 10^3-10^5\ M_\odot$ (SBS 0335-052E)
$\sim 10^4-10^5\ M_\odot$ (Tol 1214 and J1205),
$\sim 10^5-10^{6.5}\ M_\odot$ (J0344),
and
$\sim 10^5\ M_\odot$ (ID 6355).
Our results suggest that these five dwarf galaxies may harbor IMBHs with masses of 
$M_{\rm BH} \sim 10^3-10^6\ M_\odot$.

\end{enumerate}

We thank Mitsuru Kokubo, Akio K. Inoue, Mami Machida, Hirofumi Noda, Hidenobu Yajima, Jeong-Gyu Kim, Ken Osuga, Kazuhiro Hada, Hiroshi Nagai, Yoshihisa Asada, Misaki Mizumoto, Taiki Kawamuro, Takuya Mushano, Kohei Ichikawa, Makoto Ando,Takato Tokuno, and Yuki Kambara for having useful discussions.

This research is based in part on data gathered with the 10-meter Keck Telescope located at W. M. Keck Observatory.
We thank the observatory personnel for help with the observations.
This paper is supported by World Premier International Research Center Initiative (WPI Initiative), MEXT, Japan, as well as the joint research program of the Institute of Cosmic Ray Research (ICRR), the University of Tokyo. This work is supported by KAKENHI 
(19H00697, 20H00180, 21H04467, JP20K22373, and 21K03622)
Grant-in-Aid for Scientific Research through the Japan Society for the Promotion of Science.
Y.I. is supported by JSPS KAKENHI Grant No. 21J20785.
This research is supported by a grant from the Hayakawa Satio Fund awarded by the Astronomical Society of Japan.
Numerical computations were in part carried out on Small Parallel Computers at Center for Computational Astrophysics, National Astronomical Observatory of Japan.
This publication makes use of data products from the Wide-field Infrared Survey Explorer, which is a joint project of the University of California, Los Angeles, and the Jet Propulsion Laboratory/California Institute of Technology, funded by the National Aeronautics and Space Administration.

\bibliography{empghatano}
\end{document}